\documentclass[sigconf]{acmart}

\usepackage{eso-pic}
\usepackage{graphicx}
\usepackage{tikz}
\usepackage{pgf}
\usepackage{lscape}
\usepackage{amssymb}
\usepackage{fancyvrb}
\usepackage{amsfonts}
\usepackage{verbatim}
\usepackage{xfrac}
\usepackage{algorithmic}
\usepackage{url}
\usepackage{amsmath}
\usepackage{multirow}
\usepackage{listings}
\usepackage{xcolor}
\usepackage{algorithm}
\usepackage{bm}
\usepackage{proof}
\usepackage{epstopdf}
\usepackage{subfig}
\usetikzlibrary{arrows,automata}
\usetikzlibrary{shapes,snakes}
\usepackage{colortbl}% http://ctan.org/pkg/xcolor

\graphicspath{{figure/}{figures/}}

\usepackage[utf8]{inputenc}
\usepackage{pgfplots}
\usepackage{amsmath}
\usetikzlibrary{patterns}

\copyrightyear{2020} 
\acmYear{2020} 
\setcopyright{acmcopyright}
\acmConference[MobiHoc '20]{ACM International Symposium on Theory, Algorithmic Foundations, and Protocol Design for Mobile Networks and Mobile Computing}{October 11 -- 14, 2020}{Online}
\acmBooktitle{ACM International Symposium on Theory, Algorithmic Foundations, and Protocol Design for Mobile Networks and Mobile Computing  (MobiHoc '20), October 11 -- 14, 2020, Online}
\acmPrice{15.00}
\acmDOI{10.1234/1234567.8901234}
\acmISBN{123-4-5678-90-1/23/45}

\begin{document}
\title[]{\textit{PolymoRF:} Polymorphic Wireless Receivers \\Through Physical-Layer Deep Learning}
% \titlenote{Produces the permission block, and copyright information}
% \subtitle{Extended Abstract}

% \author{Francesco Restuccia and Tommaso Melodia} 
% % \authornote{Note}
% % \orcid{1234-5678-9012}
% \affiliation{%
%   \institution{Institute for the Wireless Internet of Things \\Northeastern University, Boston, MA, USA}
% }
% \email{{frestuc,  melodia}@northeastern.edu}

\author{Francesco Restuccia}
% \authornote{Note}
\affiliation{%
  \institution{Institute for the Wireless Internet of Things\\Northeastern University, Boston, MA USA}
  \streetaddress{360 Huntington Ave}
  \postcode{02115}
}
\email{frestuc@northeastern.edu}

\author{Tommaso Melodia}
% \authornote{Note}
\affiliation{%
  \institution{Institute for the Wireless Internet of Things\\Northeastern University, Boston MA USA}
  \streetaddress{360 Huntington Ave}
  \postcode{02115}
}
\email{melodia@northeastern.edu}

\renewcommand{\shortauthors}{F.~Restuccia and T.~Melodia}

\begin{abstract}
Today's wireless technologies are largely based on inflexible designs, which makes them inefficient and prone to a variety of wireless attacks. To address this key issue, wireless receivers will need to (i) infer on-the-fly the physical-layer parameters currently used by transmitters; and if needed, (ii) change their hardware and software structures to demodulate the incoming waveform.  In this paper, we introduce \emph{PolymoRF}, a deep learning-based polymorphic receiver able to reconfigure itself in real time based on the inferred waveform parameters. Our key technical innovations are (i) a novel embedded deep learning architecture, called \emph{RFNet}, which enables the solution of key waveform inference problems; (ii) a generalized hardware/software architecture that integrates \emph{RFNet} with radio components and signal processing. We prototype  \emph{PolymoRF} on a custom software-defined radio platform, and show through extensive over-the-air experiments that (i) \emph{RFNet} achieves similar accuracy to that of state-of-the-art yet with 52x and 8x latency and hardware reduction; (ii) \textit{PolymoRF} achieves throughput within 87\% of a perfect-knowledge \emph{Oracle} system,  thus demonstrating for the first time that polymorphic receivers are feasible and effective. 
\end{abstract}



 
 %Although cognitive radios have been envisioned 20 years ago, literature still lacks of practical demonstrations of polymorphic receivers.

\maketitle

\section{Introduction}

It has been forecast that over 50 billion mobile devices will be soon connected to the Internet, creating the biggest network the world has ever seen \cite{CiscoEstimates,EricssonMobility2018}.~However, only very recently has the community started to acknowledge that squeezing billions of devices into tiny spectrum portions will inevitably create disruptive levels of interference \cite{pena2017internet}. Although Mitola and Maguire first envisioned the concept of ``cognitive radios'' 20 years ago \cite{mitola1999cognitive}, today's commercial wireless devices still use inflexible wireless standards such as WiFi and Bluetooth -- and thus, are still very far from being truly real-time reconfigurable. Just to give an example of the seriousness of the spectrum inflexibility issue, DARPA has recently invested to launch the spectrum collaboration challenge (SC2), where the target is to design spectrum access schemes that ``[...] best share spectrum with any network(s), in any environment, without prior knowledge, leveraging on machine-learning techniques'' \cite{DARPA-SC2,Yu-jsac2019}.  

From a security perspective, another key issue is perhaps even more worrisome. It has been extensively demonstrated that jamming strategies targeting the inflexibility of key components of the wireless transmission, such as headers and pilots, can significantly decrease the system throughput while increasing the jammer stealthiness. For example, Clancy \cite{ClancyICC11} demonstrated that pilot nulling attacks in OFDM systems can be up to 7.5dB more effective than traditional jamming. Moreover, Vo \emph{et al.} \cite{Vo-wisec2016} show that short bursts across carefully-selected WiFi sub-carriers can destroy over 95\% of WiFi transmissions with an energy cost three orders of magnitude less than the communicating nodes.  

\begin{figure}[!h]
  \centering
  \includegraphics[width=\linewidth]{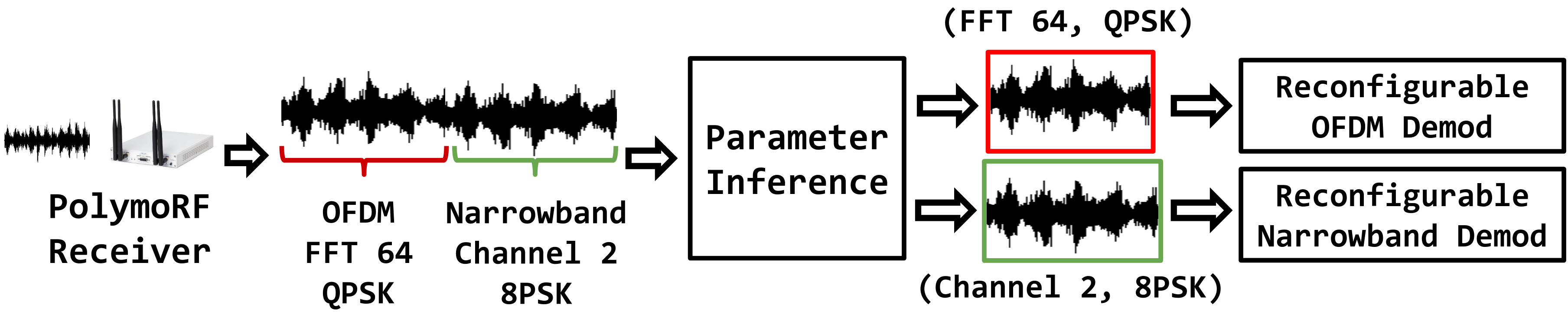}
   \caption{\label{fig:poly_system}Example of a Self-Adaptive Polymorphic Receiver.}\vspace{-0.2cm}
\end{figure}

Intuitively, the issues of existing communication systems could be addressed by allowing transmitters to dynamically switch parameters such as carrier frequency, FFT size, and symbol modulation without coordination with the receiver. This will allow the transmitter (i) efficient spectrum occupation by using the most appropriate wireless scheme at any given moment, and (ii) change position of header and pilots over time and thus becoming less jamming-prone. Figure \ref{fig:poly_system} shows an example of a polymorphic receiver able to infer the current transmitter's physical-layer scheme (\textit{e.g.}, OFDM vs narrowband) and the scheme's parameters (\textit{e.g.}, FFT size, channel, modulation), and then demodulate each portion of the signal. 

% De Bruhl \emph{et al.} \cite{debruhl2013jam} have experimentally proved that through careful selection of jamming period and duration, a jammer with a 4\% duty cycle is able to lower the IEEE 802.15.4 packet delivery ratio to only 5\%

Doing away with explicit coordination and inflexible physical layers is the first step toward wireless receivers able to self-adapt to demodulate many waveform with a single radio interface \cite{Restuccia-IoT2018,restuccia2020physical}. Yet, despite their compelling necessity, these wireless receivers do not exist today. This manuscript aims to change the current state of affairs by proposing the first demonstration of \emph{PolymoRF}, the first \textit{polymorphic wireless receiver}. Achieving this goal required us to address a set of key research challenges summarized below: \smallskip 

%In our vision, polymorphism will concretely realize a \emph{minimalistic, protocol-free, inference-based, on-the-fly} approach to wireless communications, where transmitters and receivers operate using a very limited set of basic rules (\textit{e.g.}, set of modulations, set of channels) which are seamlessly changed by the transmitter at will, possibly \textit{without using control channels or headers}. The receiver, in turn, infers the parameters using machine learning techniques, and then ``morphs'' itself into a new configuration to demodulate the data.

\textbf{(1)} \emph{Keeping Up with the Transmitter.}~A crucial aspect is the real-time parameter inference.~In practical systems, however, transmitters may choose to switch its parameter configuration in the order of milliseconds (\textit{e.g}., frequency hopping, rate adaptation). For example, if the transmitter chooses to switch modulation every 100ms, the learning model should run in (much) less than 100ms (as computed in Section \ref{sec:par_strategies}) to predict the parameters and morph the receiver into a new configuration. To this end, we will show in Section \ref{sec:hardware_eval} that CPU latency is several orders of magnitude greater than what is required to sustain realistic sampling rates from the RF interface. Thus, we need hardware-based designs to implement low-latency knowledge extraction techniques. \smallskip

\textbf{(2)} \emph{Creating Learning Architectures for the Embedded RF Domain.}  Recent advances in RF deep learning \cite{Kulin-ieeeaccess2018,OShea-ieeejstsp2018,Karra-ieeedyspan2017,o2017introduction,o2016convolutional,restuccia2020deepwierl} have demonstrated that convolutional neural networks (ConvNets) may be applied to analyze RF data without feature extraction and selection algorithms \cite{Xu-ieeetvt2010,Pawar-ieeetifs2011,Shi-ieeetcomm2012,Ghodeshar-icscn2015}. Moreover, ConvNets present a number of characteristics (discussed in Section \ref{sec:rf_images}) that make them particularly desirable from a hardware implementation perspective. However, these solutions cannot be applied to implement real-time polymorphic wireless communications--as shown in Section \ref{sec:hardware_eval}, existing art \cite{OShea-ieeejstsp2018,o2016convolutional} utilizes general-purpose architectures with a very high number of parameters, requiring hardware resources and latency that go beyond what is acceptable in the embedded domain. This crucial issue calls for novel, RF-specific, real-time architectures. We are not aware of learning systems tested in a real-time wireless environment and used to implement inference-based wireless systems.\smallskip

\textbf{(3)} \emph{System-level Feasibility of Polymorphic Platforms.}~It is yet to be demonstrated whether polymorphic platforms are feasible and effective. This is not without a reason -- from a system perspective, it required us to tightly interconnect traditionally separated components, such as CPU, RF front-end, and embedded operating system/kernel, to form a seamlessly-running low-latency learning architecture closely interacting with the RF components and able to adapt at will its hardware and software based on RF-based inference.~Furthermore, since polymorphic wireless systems are subject to inference errors, we need to test its performance against a perfect-knowledge (thus, ideal and not implementable) system.   \vspace{-0.2cm}

%Second, perhaps even more importantly, \textit{RF signals are subject to unpredictable impairments given by the RF channel}, which implies the input data will be non-stationary \cite{goodfellow2016deep}. Thus, we need to investigate whether ConvNets can be made robust to channel impairments.

\subsection*{Technical Contributions}

%We believe that the above core challenges can only be addressed through novel, IoT-specific DRL designs and architectures. As such, this paper proposes \textit{Deep Wireless Embedded Reinforcement Learning} (\textit{DeepWiERL}), the first DRL system carefully designed to bridge the existing gap between theoretical and system-level aspects of wireless DRL in the IoT landscape. The key and critical innovation behind \emph{DeepWiERL} is to bring to the IoT research community what has been missing so far -- a  general-purpose framework to design, implement and evaluate the performance of IoT-tailored real-time DRL algorithms on embedded devices.

This paper's key innovation is to finally bridge the gap between the extensive theoretical research on cognitive radios and the associated system-level challenges, by demonstrating that inference-based wireless communications are indeed feasible on off-the-shelf embedded devices. Beyond the examples and the evaluation conducted in Section \ref{sec:exp_results}, the main purpose of this work is to provide a \textit{blueprint} for next-generation wireless receivers, where their radio hardware and software are not \textit{protocol-specific}, but instead \textit{spectrum-driven} and adaptable on-the-fly to different waveforms. \vspace{0.1cm}

We summarize our main technical contributions as follows: \smallskip

\textbf{(1)} We design a novel learning architecture called \textit{RFNet}, specifically and carefully tailored for the embedded RF domain. Our key intuition in \textit{RFNet} is to arrange I/Q samples to form an ``image'' that can be effectively analyzed by the ConvNet filters. This operation produces high-dimensional representations of small-scale transition in the I/Q complex plane, which can be leveraged to efficiently solve a wide variety of complex RF classification problems such as RF modulation classification. Extensive experimental evaluation indicates that \textit{RFNet} obtains similar accuracy achieved by prior art \cite{OShea-ieeejstsp2018,o2016convolutional}, while reducing latency and hardware by 52x and 8x;\smallskip

\textbf{(2)} We propose a general-purpose hardware/software architecture for software-defined radios that enables the creation of custom polymorphic wireless systems through \emph{RFNet}. Then, we implement a multi-purpose library based on high-level synthesis (HLS) that translates an \emph{RFNet} model implemented in software to a circuit implemented in the FPGA portion of the SoC. Moreover, we leverage key optimization strategies such as pipelining and unrolling to further reduce the latency of \emph{RFNet} by more than 50\% with respect to the unoptimized version, with only 7\% increase of hardware resource consumption. Finally, we design and implement the device-tree entries and Linux drivers enabling the system to utilize \emph{RFNet} and other key hardware peripherals; \smallskip

\textbf{(3)} We prototype \emph{PolymoRF} on a ZYNQ-7000 system-on-chip (SoC) and analyze its performance on a scheme where the transmitter can switch among 3 FFT sizes and 3 symbol modulation schemes without explicit notification to the receiver. A demo video of PolymoRF where the transmitter switches FFT size every 0.5s is available at \url{https://youtu.be/5vf_pb0nvKk}. We believe ours is the first demonstration of real-time OFDM reconfigurability without explicit transmitter/receiver coordination. Experiments on both line-of-sight (LOS) and non-line-of-sight (NLOS) channel conditions show that the system achieves at least  87\% of the throughput of a perfect-knowledge -- and thus, unrealistic -- \emph{Oracle} OFDM system, thus proving the feasibility of polymorphic receivers.

\section{\texorpdfstring{P\MakeLowercase{olymor}RF}{PolymoRF}: An Overview} \label{sec:system}

\begin{figure}[!h]
  \centering
  \includegraphics[width=\linewidth]{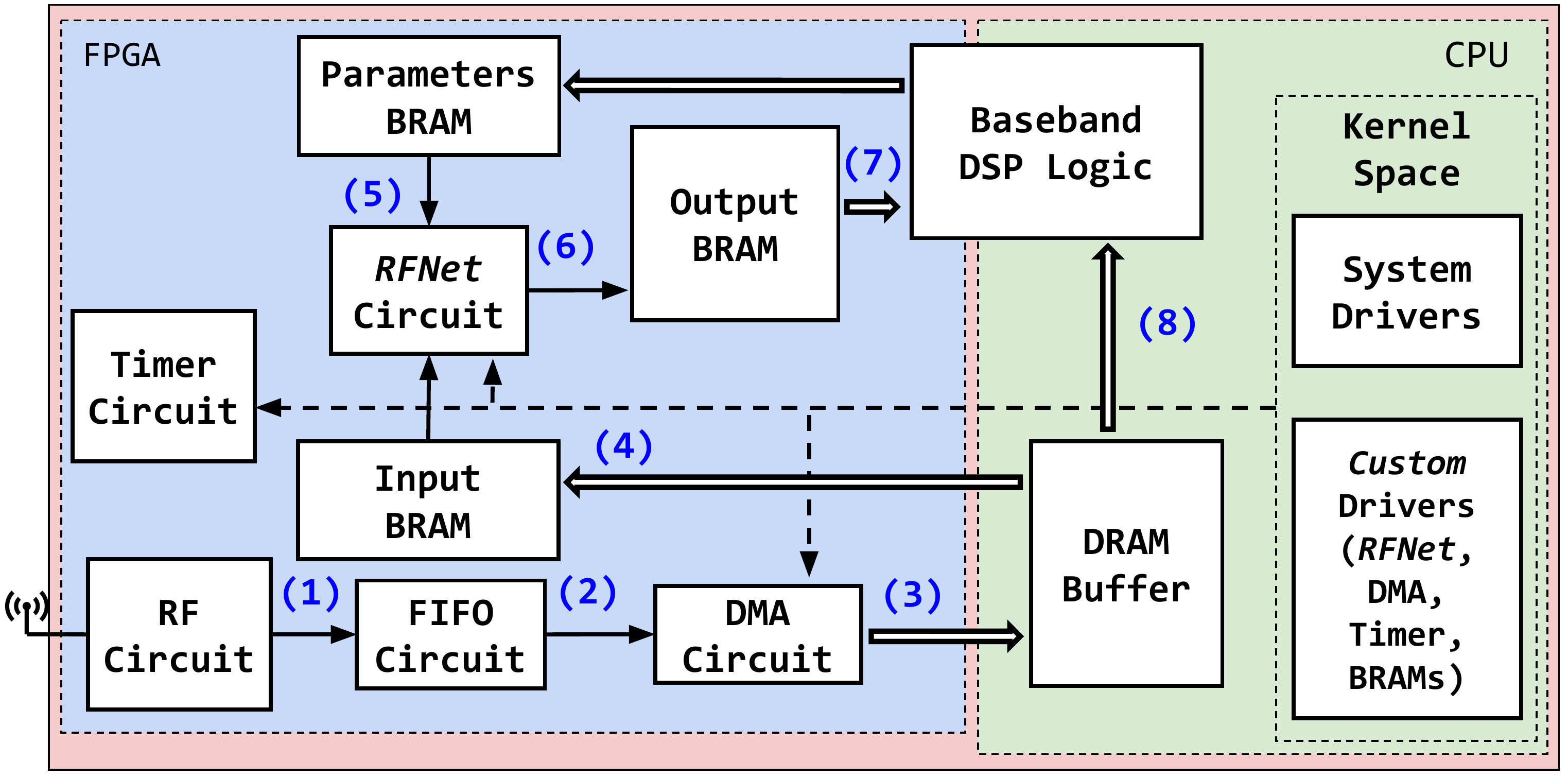}
   \caption{\label{fig:arch}Modules and Operations of \emph{PolymoRF}.\vspace{-0.3cm}}
\end{figure}

The primary operations performed by the \emph{PolymoRF} platform are summarized in Figure \ref{fig:arch}. In a nutshell, \emph{PolymoRF} can be considered as a full-fledged learning-based software-defined radio architecture where both the inference system and the demodulation strategy can be morphed into new configurations at will.

We provide a walk-through of the main operations performed by \emph{PolymoRF} with the help of Figure \ref{fig:arch}. Although for simplicity we refer to specific hardware equipment and circuits in our explanation, we point out that the building blocks of our platform design (BRAMs, DMA, FIFOs, etc) can be implemented in any commercially-available FPGA platform. 

We assume the transmitter may transmit by choosing among a discrete set of physical-layer parameters which are known at the receiver's side.  We define as $Y$ a tuple of such physical-layer parameters, which may be changed at will by the transmitter but not before $T_{sw}$ seconds between each change, which we refer to a \emph{switching time}. For the sake of generality, in this paper we will not assume any particular strategy in the transmitter's parameter choice, which can be driven by a series of factors (including anti-jamming strategy, noise avoidance, throughput optimization, and so on) that will be considered as out of the scope of this paper, whose main focus is instead on the receiver's side.\vspace{0.1cm}

% \begin{figure}[!h]
%     \centering
%     \includegraphics[width=\linewidth]{./figures/Mobicom_ZC706.pdf}
%     \caption{Prototype testbed for \emph{PolymoRF}.}
%     \label{fig:zc706}
%     \vspace{-0.3cm}
% \end{figure}

\textbf{(1)}~\emph{Reconfigurable Radio Front-end.}~The RF signal is received (step 1) through a reconfigurable RF front-end. In our prototype, we used an AD9361 \cite{AD9361} radio interface, which supports frequency range between 70 MHz to 6.0 GHz and channel bandwidth between 200 kHz to 56 MHz. We chose the AD9361 because it is commonly used in software-defined radio systems -- indeed, it is also used by USRPs such as the E310 and B210. Moreover, the AD9361 provides basic FPGA reference designs and kernel-space drivers to ease prototyping and extensions. Perhaps more importantly, the AD9361 local oscillator (LO) frequency and RF bandwidth can be reconfigured at will through CPU registers. \smallskip

\textbf{(2)}~\emph{Conversion from RF to FPGA domain.}~The AD9361 produces streams of I/Q samples of 200M samples/second -- hence, it is clocked at 200 MHz. Since the AD9361 clock would be too fast for the other circuits in the FPGA, we implemented a FIFO to adapt the speed of samples from the AD9361 to the 100 MHz clock frequency used by the other circuits in the FPGA (step 2). We then use a direct memory access (DMA) core to store the stream of I/Q samples to a buffer in the DRAM (step 3). The use of DMA is crucial as the CPU cannot do the transfer itself, since it would be fully occupied for the entire duration of the read/write operation, and thus unavailable to perform other work. Therefore, we wrote a custom DMA driver to periodically fill a buffer of size $B$ residing in the DRAM with a subset of I/Q samples coming from the FIFO.\smallskip

\textbf{(3)}~\emph{Learning and Receiver Polymorphism.}~After the buffer has been replenished, its first $X$ I/Q samples are sent to a BRAM (step 4) constituting the input to \textit{RFNet}, a novel learning architecture based on \textit{ConvNets}. This circuit is the fundamental core of the \textit{PolymoRF} system; therefore, we will dedicate Sections \ref{sec:rf_images} and \ref{sec:soc_implementation} to discuss in details its architecture and implementation, respectively. The parameters of \emph{RFNet} are read by an additional BRAM (step 5), which in effect allows the reconfiguration of \emph{RFNet} to address multiple RF problems according to the current platform need. As explained in Section \ref{sec:rf_images}, \emph{RFNet} produces a probability distribution over the transmitter's parameter set $Y$. After \emph{RFNet} has inferred the transmitter's parameters, it writes on a block-RAM (BRAM) its probability distribution (step 6). Then, the baseband DSP logic (which may be implemented in both hardware and software) reads the distribution from the BRAM (step 7), selects the parameter set with highest probability, and ``morphs'' into a new configuration to demodulate the I/Q samples in $B$ (step 8).

\section{Learning System: \texorpdfstring{RFN\MakeLowercase{et}}{RFNet}}\label{sec:rf_images}

We first motivate the use of convolutional neural networks for RFNet, then we discuss some RF-specific learning challenges, and then we describe in details the RFNet input construction and its complete architecture.\smallskip

\textbf{(1)}~\emph{Why Using Deep Learning and not Machine Learning?}~Deep learning relieves from the burden of finding the right ``features'' characterizing a given wireless phenomenon.~At the physical layer, this is a key advantage for the following reasons. First, deep learning offers high-dimensional feature spaces.~In particular, O'Shea \emph{et al.} \cite{OShea-ieeejstsp2018} have demonstrated that on the 24-modulation dataset considered, deep learning models achieve on the average about 20\% higher classification accuracy  than legacy learning models under noisy channel conditions. Second, automatic feature extraction allows to reuse the same hardware circuit to address different learning problems. Critically, this allows to keep both latency and energy consumption constant, which are particularly critical in wireless systems. Third, deep learning algorithms can be fine-tuned by performing batch gradient descent on fresh input data, avoiding manual re-tuning of the feature extraction algorithms.\smallskip

\textbf{(1)}~\emph{Why Using ConvNets for Wireless Deep Learning?}~There are several primary advantages that make the usage of ConvNet-based models particularly desirable for the embedded RF domain. First, \textit{convolutional filters are designed to interact only with a very small portion of the input}. We show in Section \ref{sec:learning_eval} that this key property allows to achieve significantly higher accuracy than traditional neural networks. Perhaps even more importantly, \textit{ConvNets are scalable with the input size}. For example, for a 200x200 input and a DL with 10 neurons, a traditional neural network will have $200^2 \cdot 10$ = 400k weights, which implies a memory occupation of $4 \cdot 400$k = 16 Mbytes to store the weights of a single layer (\textit{i.e.}, a float number for each weight). Clearly, this is unacceptable for the embedded domain, as the network memory consumption would become intractable as soon as several DLs are stacked on top of the other. 

Moreover, as we show in Section \ref{sec:par_strategies}, \textit{ConvNet filtering operations can be made low-latency by parallelization}, which makes them particularly suitable to be optimized for the RF domain. Finally, we show in Section \ref{sec:exp_results} that the same ConvNet architectures can be reused to address different RF classification problems (\textit{e.g.}, modulation classification in single- and multi-carrier systems), as long as the ConvNet is provided appropriate weights through training. Our ConvNet hardware design (Section \ref{sec:learning_rf_circuits}) has been specifically designed to allow seamless ConvNet reconfiguration and thus solving different RF problems according to the system's needs.\vspace{0.1cm}

\begin{figure}[!h]
    \centering
    \includegraphics[width=\columnwidth]{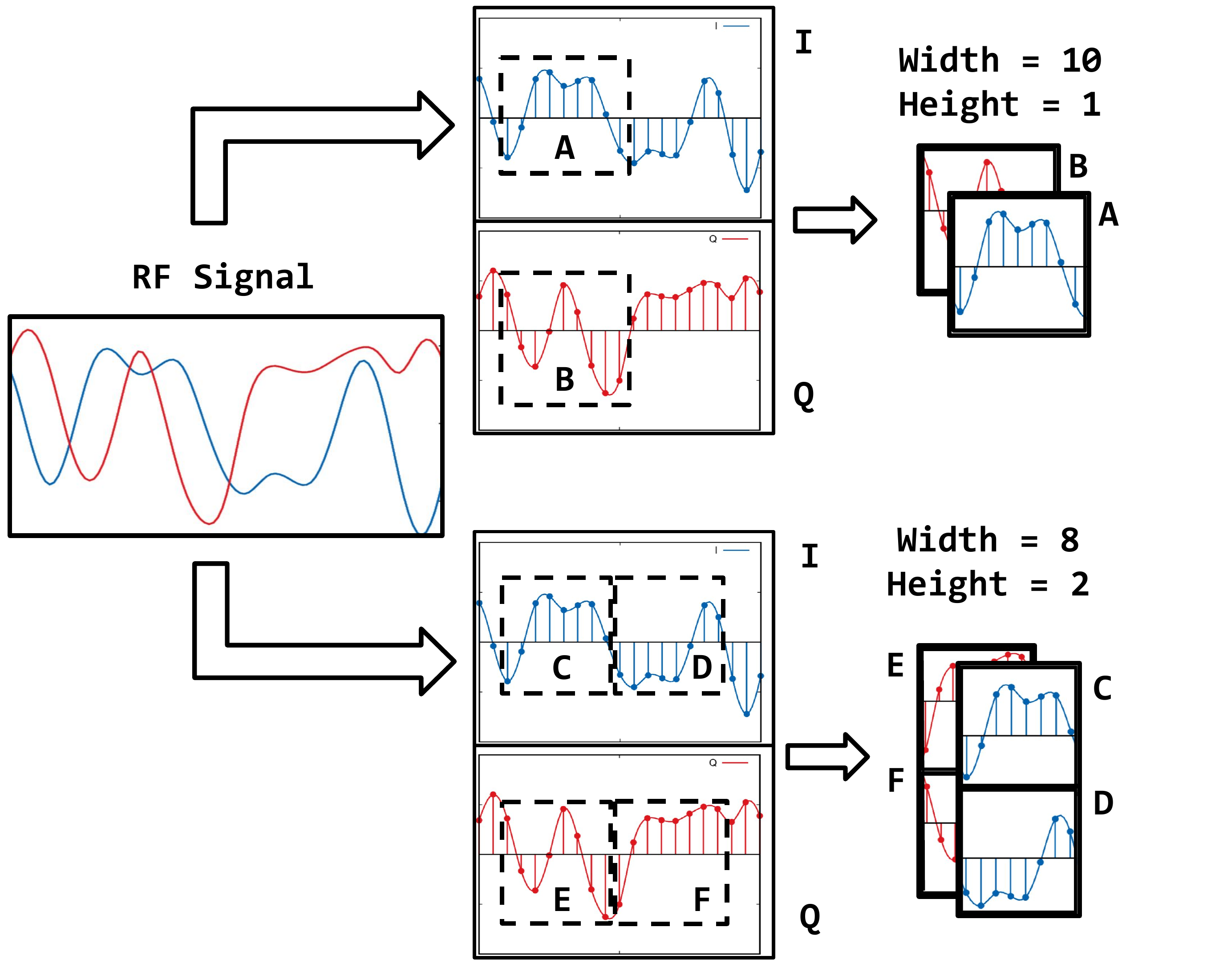}
    \caption{How \emph{RFNet} constructs tensors from I/Q samples.}
    \label{fig:rf_signal_image}\vspace{-0.4cm}
\end{figure}

\textbf{(2)}~\emph{RF-specific Learning Challenges.} There are a number of key challenges in RF learning that are substantially absent in the CV domain. Among others, we know that RF signals are continuously subject to dynamic (and usually unpredictable) noise/interference coming from various sources. This may decrease the accuracy of the learning model. For example, portions of a QPSK transmission could be mistaken for 8PSK transmissions since they share part of their constellations. We address the above core design issues with the following intuitions. First, although RF signals are affected by fading/noise, in most practical cases their effect can be considered as constant over small intervals. Second, though some constellations are similar to each other, the transitions between the symbols of the constellations are distinguishable when the waveform is sampled at a higher sampling rate than the one used by the transmitter. Third, convolution operations are equivariant to translation, so they can recognize I/Q patterns regardless of where they occur. \vspace{0.1cm}

\begin{figure}[!h]
    \centering
    \includegraphics[width=\columnwidth]{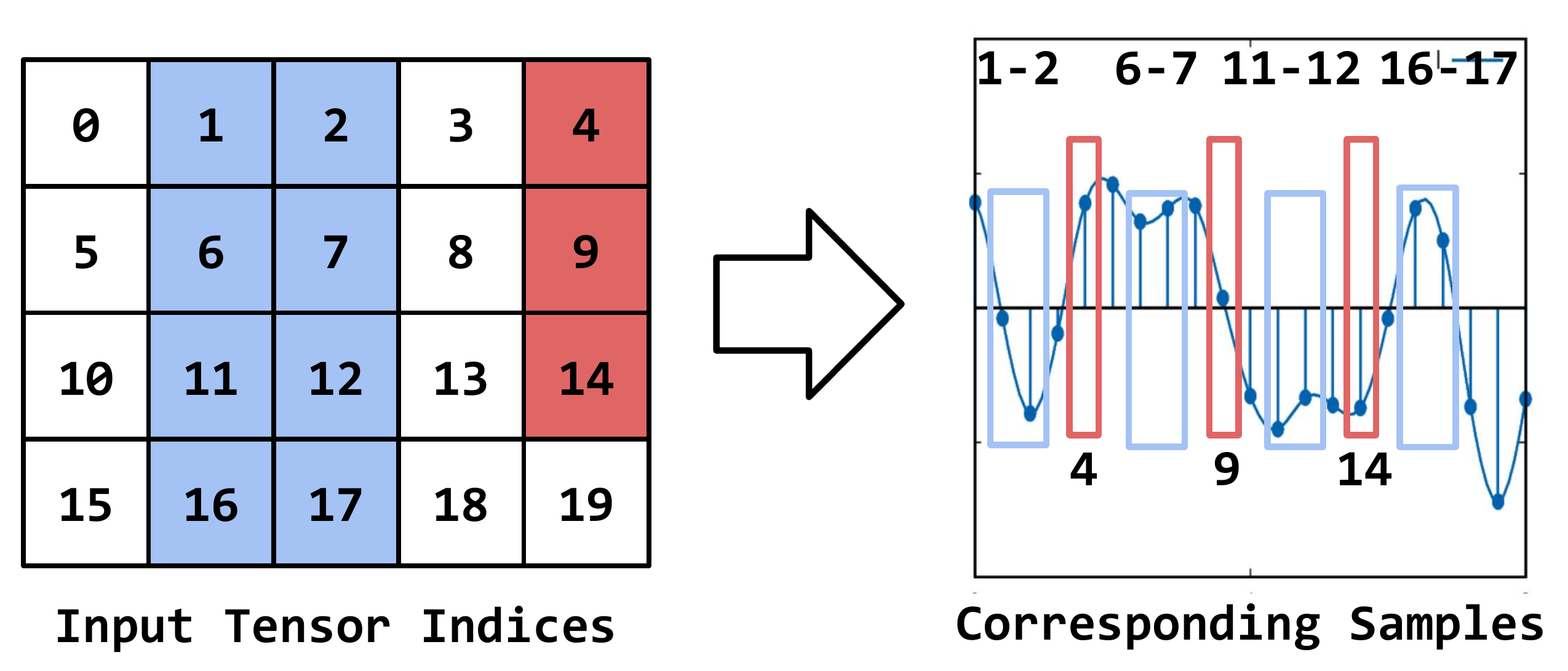}
    \caption{\emph{RFNet} captures small-scale I/Q pattern sequences.}
    \label{fig:periodic_filter}
\end{figure}

%This is fundamental in the RF domain, as I/Q patterns in the signal data can be present in any portion of the RF signal and are usually function of a small number of neighboring I/Q samples. 

%Finally,  Another key challenge in applying ConvNet to the RF domain is transforming the RF waveforms into an ``image'' (\textit{i.e.}, input tensor) representation that a ConvNet can easily manipulate through different transformations -- and thus, learn its patterns as done in the CV domain.  Such input tensor must be carefully designed to (i) facilitate the learning of unique RF-based features by the filters in the convolutional layers; and (ii) increase the ConvNet's resilience to RF noise.

\textbf{(3)}~\emph{RFNet Input Construction.} By leveraging these key concepts, we can design a learning system that distinguishes waveforms by recognizing  transitions in the  I/Q complex plane regardless of where they happen, by leveraging the shift-invariance property of convolutional layers. More formally, let us consider a discrete-time complex-valued I/Q sequence $s[k]$, where $k \ge 0$. Let us consider $M = W \cdot H$ consecutive I/Q samples $s[j],\ 0 \le j \le W\cdot H$, where $W$ and $H$ are the \textit{width} and \textit{height} of the input tensor. The input tensor $\mathcal{T}$, of dimension $W\times H \times 2$, is constructed as follows: \vspace{-0.3cm}

%Usually, RF waveforms are represented as a discrete sequence of real-valued samples, each corresponding to the amplitude of the waveform at a given moment in time. However, in this paper we will use the I/Q representation of the RF signal to create \emph{PolymoRF}'s input tensors. We make this choice for the following reasons: (i) off-the-shelf RF circuitry already provides I/Q representation of RF waveforms, so the I/Q representation of an RF waveform comes at no additional cost; (ii) ConvNet natively support multi-dimensional data by allowing input tensors of arbitrary number of dimensions; and (iii) most importantly, since digital modulation schemes manipulate the I/Q components of the RF waveform in a separate way, I/Q components will significantly aid the learning process of our ConvNet models. 

% We point out that our choice of input tensor representation is not casual, but indeed reflects a precise and careful choice in representing I/Q data so that convolutional layers are facilitated to learn RF data. \textbf{Our key intuition is that the width and height dimension of the input tensor together determine the amount of both periodic and aperiodic RF waveform characteristics that we want the filters in the convolutional layers to capture}.~Indeed, 

\begin{eqnarray}\nonumber
    \mathcal{T}[r, c, d]  =  \mbox{Re}\left\{s[r \cdot W + c]\right\} \cdot (1 - d) +\\
     \mbox{Im}\{s[r \cdot W + c]\}\cdot d,\\ \nonumber
     \mbox{ where } d\in \{0, 1\}, 0 \le r \le H, 0 \le c \le W
\end{eqnarray}

By construction, it follows that $\mathcal{T}[r+1, c] = s[(r + 1) \cdot W + c] = s[r \cdot W + c + W]$, meaning that (i) I/Q samples in adjacent columns will be spaced in time by a factor of 1, and (ii) I/Q samples in adjacent rows will be spaced in time by a factor of $W$; moreover, (iii) our input tensors have depth equal to 2, corresponding to the I and Q data, respectively, which will allow the \textit{RFNet} filters to examine each element of the input tensor \textit{without decoupling the I and Q components of the RF waveform}. Figure \ref{fig:rf_signal_image} depicts an example of a 2x4 and 1x3 filters operating on a waveform. 

\begin{figure}[!h]
    \centering
    \includegraphics[width=\columnwidth]{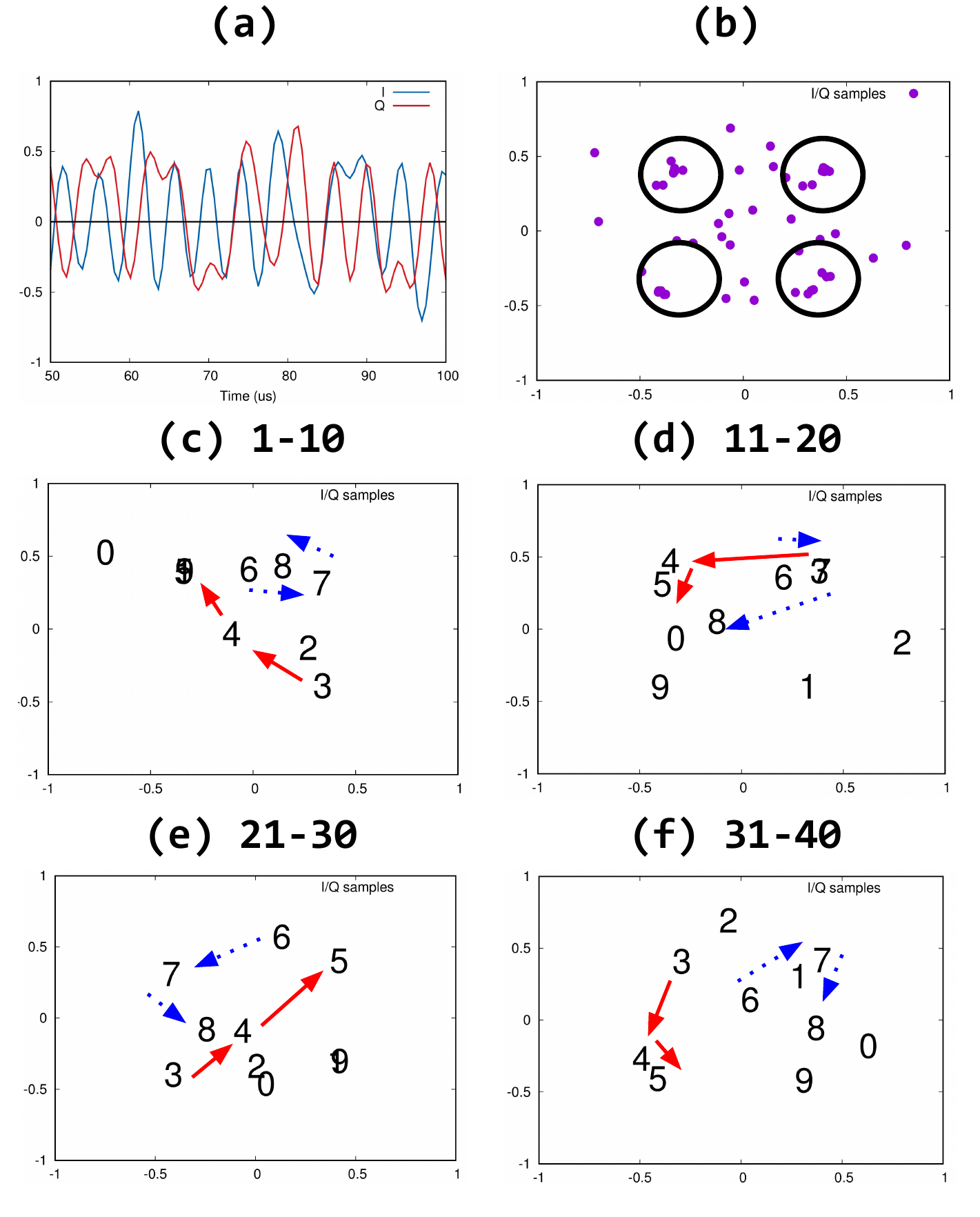}
    \caption{Example of I/Q patterns in a QPSK waveform.\vspace{-0.2cm}}
    \label{fig:filters_intuition}
\end{figure}

To further show the intuition behind our architectural design, Figures \ref{fig:filters_intuition}(c)-(f) show the first 40 I/Q samples (ordered by occurrence) of a QPSK waveform in  \ref{fig:filters_intuition}(a) which corresponds to the I/Q constellation shown in  \ref{fig:filters_intuition}(b). We consider an example where $3\times  3$ filters are used, with $W = 10$ and $H = 4$. \textit{By setting $W$ equal to the sampling rate, we ``force'' the consecutive rows of the filter to recognize transitions occurring in the same portion of the symbol transmitted by the receiver.} To make this point, we show in red and blue arrows the I/Q transitions ``seen'' by a filter applied at indices 3-4-5 and 6-7-8, respectively. While in the former case the transitions between the I/Q constellation points peculiar to QPSK can be clearly recognized, in the latter case the filter will not recognize patterns. This is because samples 3-4-5 are in between QPSK constellation points, while 7-8-9 are close to   QPSK. 

\begin{figure}[!h] 
    \centering
    \includegraphics[width=\linewidth]{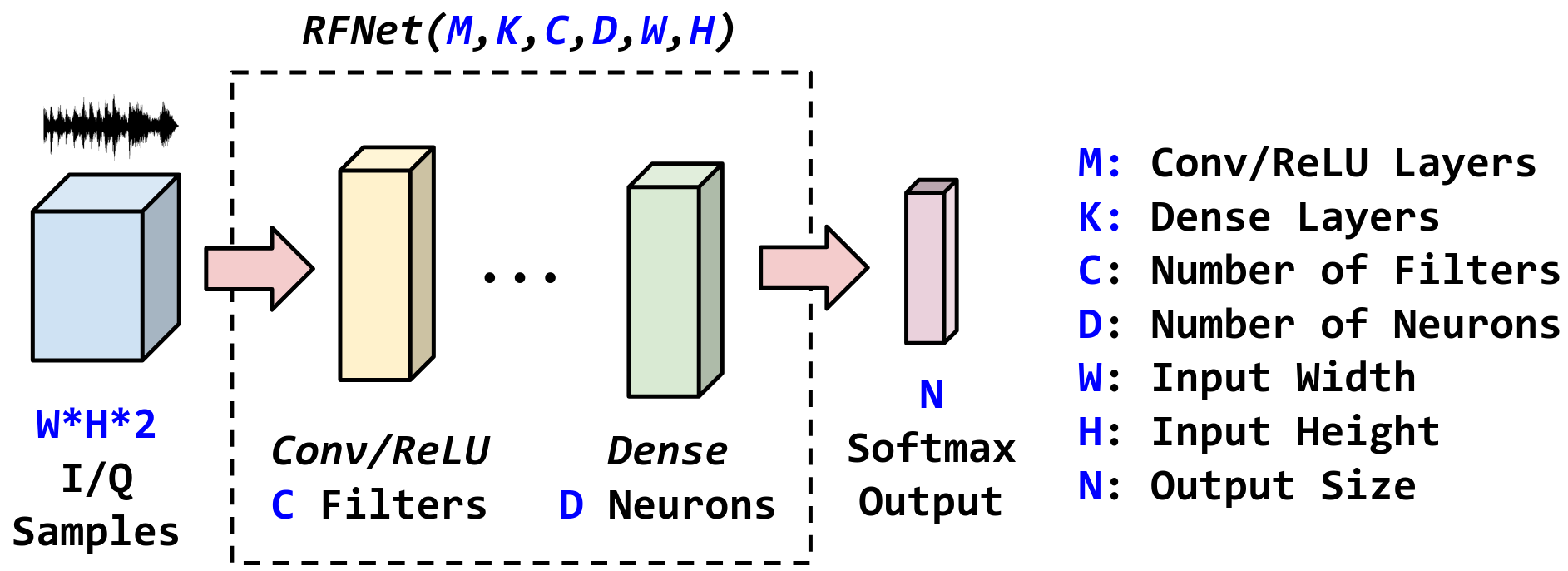}
    \caption{RFNet Architecture.}
    \label{fig:rfnet}
    \vspace{-0.3cm}
\end{figure}

Figure \ref{fig:rfnet} shows the complete architecture of RFNet. Similar to existing work \cite{OShea-ieeejstsp2018} and computer vision-based models, the network is composed by $M$ convolutional (Conv) layers with $C$ filters each, followed by rectified linear units (ReLU) as activation functions. These are then followed by a series of dense layers, each having $D$ neurons. The final layer is a softmax output, which gives the probability distribution over the set of all possible classes.

% % From an implementation perspective, ConvNets are very efficient from both a space and time perspective. They are highly parallelizable by definition at the layer level and at the filter level; indeed, the computation of different filters maps can be done in parallel, as well as the operations to compute a single filter map (summations in Equation \ref{eq:filters}). 

\section{\texorpdfstring{P\MakeLowercase{olymor}RF}{PolymoRF}: HW/SW Architecture}\label{sec:soc_implementation}

This section presents the hardware and driver design and implementation of our \emph{PolymoRF} system. We first discuss the design, hardware implementation and main operations of \emph{RFNet} in Section \ref{sec:learning_rf_circuits}, followed by a discussion of the parallelization strategies in Section \ref{sec:par_strategies} and of the Linux drivers implemented to operate the circuits in Section \ref{sec:linux_drivers}.\vspace{-0.2cm}

%\subsection{Design Challenges and Constraints}\label{sec:design_challenges}

%HLS has several advantages over traditional approaches. First, it helps decrease the challenging programming efforts required to translate an algorithm to hardware. Second, an HLS toolchain can tell how many cycles are needed for a circuit to generate all the outputs for a given input size, given a target parallelization level. This helps \emph{PolymoRF} to make the best trade-off between hardware complexity and latency.

%We will show in Section \ref{sec:rf_images} that the same ConvNet architecture can be used to solve multiple classification problems. For example, if the transmitter's modulation is fixed and known in advance by the receiver, we can increase the ConvNet's accuracy by classifying onto a lower output space which includes center frequency and sampling rate but not modulation. For this reason, \emph{PolymoRF} supports the reconfiguration of the ConvNet parameters at run time by the CPU, which is done by writing onto a block random access memory (BRAM) in the hardware portion of the platform (step 8), so that the ConvNet can use the new parameters to classify new transmissions

\subsection{RFNet: Architecture and Operations}\label{sec:learning_rf_circuits}

\textbf{(1)}~\emph{Design Constraints.}~One the core design issues to address is ensuring that \emph{the same RFNet circuit can be reused for multiple learning problems and not just one architecture}. For example,  the wireless node might want to classify only specific properties of an RF waveform, \textit{e.g.}, classify only modulation since the FFT size is already known. This requires reconfigurability of the model parameters, as the device's hardware constraints may not be able to accommodate multiple learning architectures. In other words, we want RFNet to be able to operate with a different set of filters and weight parameters according to the circumstances. For this reason, we have used high-level synthesis (HLS) to design a library that translates a Keras-compliant \textit{RFNet} into an FPGA-compliant circuit. HLS is an automated design process that interprets an algorithmic description of a desired behavior (\textit{e.g.}, C/C++) and creates a model written in hardware description language (HDL) that can be executed by the FPGA \cite{winterstein2013high}. \vspace{0.1cm}

\begin{figure}[!h]
    \centering
    \includegraphics[width=\columnwidth]{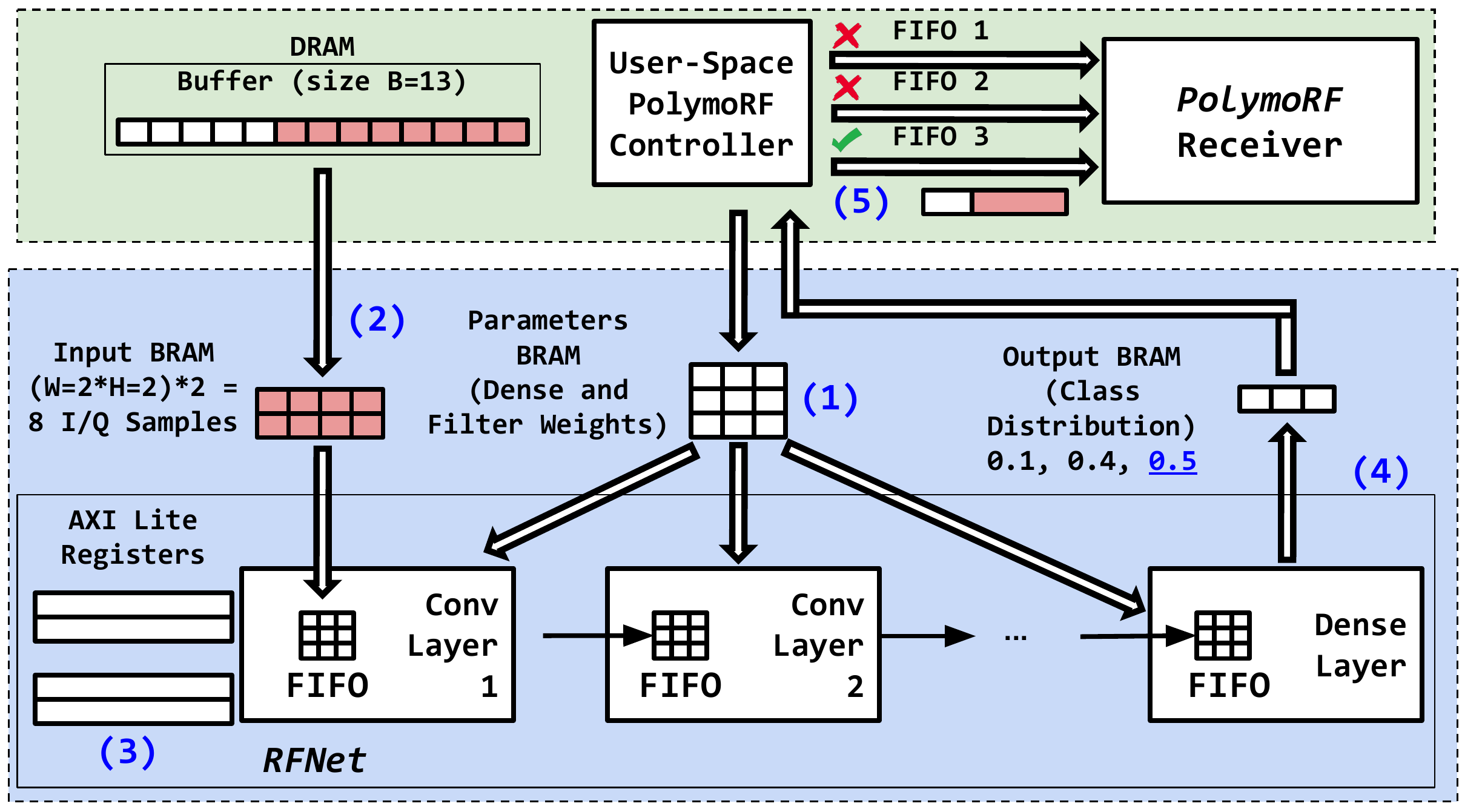}
    \caption{Block scheme of \textit{PolymoRF}'s learning circuit.}
    \label{fig:layer}
\end{figure}

\textbf{(2)}~\emph{Circuit Design.}~Figure \ref{fig:layer} shows a block scheme of our HLS-based \textit{RFNet} circuit and its main interactions with the CPU and other FPGA components. We also provide an example with some numbers to ease presentation. The main feature of our RFNet implementation is its modularity -- indeed, the circuits implementing each layer are \textit{independent from each other}, which allows for ease of parallelization and transition from HLS to HDL. Consecutive layers in \textit{RFNet} exchange data through high-speed AXI-Stream interfaces that then store the results of each layer in a FIFO, read by the next layer. Our architecture uses a 32-bit fixed-point representation for real numbers, with 10 bits dedicated to the integer portion. We chose fixed-point instead of floating-point to decrease drastically computation and hardware architecture complexity, as we do not need the precision of floating-point arithmetic. Another key advantage of our implementation is that it clearly separates the computation from the parameters, which allows for \textit{seamless real-time reconfigurability}. This is achieved by writing the parameters in a BRAM accessible by the CPU and by the \textit{RFNet} circuit. \vspace{0.1cm}

\textbf{(3)}~\emph{Main Operations.}~The first operation is to write the \textit{RFNet}'s parameters into a BRAM through the user-space PolymoRF controller (step 1). These parameters are the weights of the convolutional layer filters and the weights of the dense layers. Since we use a fixed-point architecture, each parameter is converted into fixed-point representation before being written to the BRAM. As soon as a new input buffer $B$ (of size 13 in our example) has been replenished, the controller writes the RFNet input (the first 8 I/Q samples in our example) into the input BRAM (step 2). \emph{RFNet} operations are then started by writing into an AXI-Lite register (step 3) through a customized kernel-level Linux driver. Once the results have been written in the output BRAM (step 4), \emph{RFNet} writes an acknowledgement bit into another AXI-Lite register, which signals the controller that the output is ready. Then, the controller reads the output (in our example, class 3 has the highest probability), and sends the entire buffer $B$ through a Linux FIFO to the PolymoRF receiver (step 5), which is currently implemented in Gnuradio software. The receiver has different FIFOs, each for a parameter set. Whenever a FIFO gets replenished, the part of the flowgraph corresponding to that parameter set activates and demodulates the I/Q samples contained in the buffer $B$. \textit{Notice that for efficiency reasons the receiver chains do not run when the FIFO is empty}, therefore only one receiver chain can be active at at time.\vspace{-0.2cm}

\subsection{RFNet: Latency Optimization}\label{sec:par_strategies}

We show in Section \ref{sec:hardware_eval} that the above implementation on the average reduces the latency by about 95\% with respect to a model implemented in the CPU. However, this performance may not be enough to sustain the flow of I/Q samples coming the RF interface.  We derive some formulas to explain this critical point. 

\textbf{(1)}~\emph{Why do we need to further optimize latency?}~Let us suppose that the RF interface is receiving samples at $S$ samples/sec. The AD9361 quantizes I/Q samples using a 16-bit ADC, so each I/Q sample occupies 4 bytes. Therefore, the system needs to process data with throughput $4\times S$ MB/sec to keep up with the sampling rate. To process $4\cdot S$ MB of I/Q data, \emph{PolymoRF} must do the following: (i) insert $B \times 8$ bytes into the DRAM through DMA; (ii) transfer the first $W \cdot H$ samples to the input BRAM; (iii) execute \emph{RFNet}; (iv) read the inference from the output BRAM, for a total of $\sfrac{4\cdot S}{B \cdot 8} = \sfrac{S}{B \cdot 2}$ times. More formally, by defining the above quantities as $T_{buf}$, $T_{i}$, $T_{cn}$, and $T_{o}$, it must hold that $(T_{buf} + T_i + T_{cn} + T_o)\cdot \sfrac{4\cdot S}{B \cdot 8} < 1$. Since our measurements  show that $T_{buf}, T_{i}, T_{o} \ll T_{cw}$, then $T_{cn} < {2 \cdot B}/{S}.$

Since sampling rate $S$ is usually fixed, to make the bound on $T_{cn}$ hold, we can either (i) increase buffer size $B$ or decrease \textit{RFNet} latency $T_{cn}$. Increasing the buffer size $B$ is not desirable since an increase in buffer size implies a decrease in switching time $T_{sw}$. To give a perspective of the order of magnitude of these quantities, in our experiments $S = 5$ MS/s and $T_{cn} \sim= 16$ms. This implies that the buffer size $B$ must be greater than 40,000 I/Q samples, implying that \textit{RFNet} inference must be valid for at least 40,000 I/Q samples, which correspond to a switching time $T_{sw} =$ 8ms.  

\begin{figure}[!h]
    \centering
    \includegraphics[width=\columnwidth]{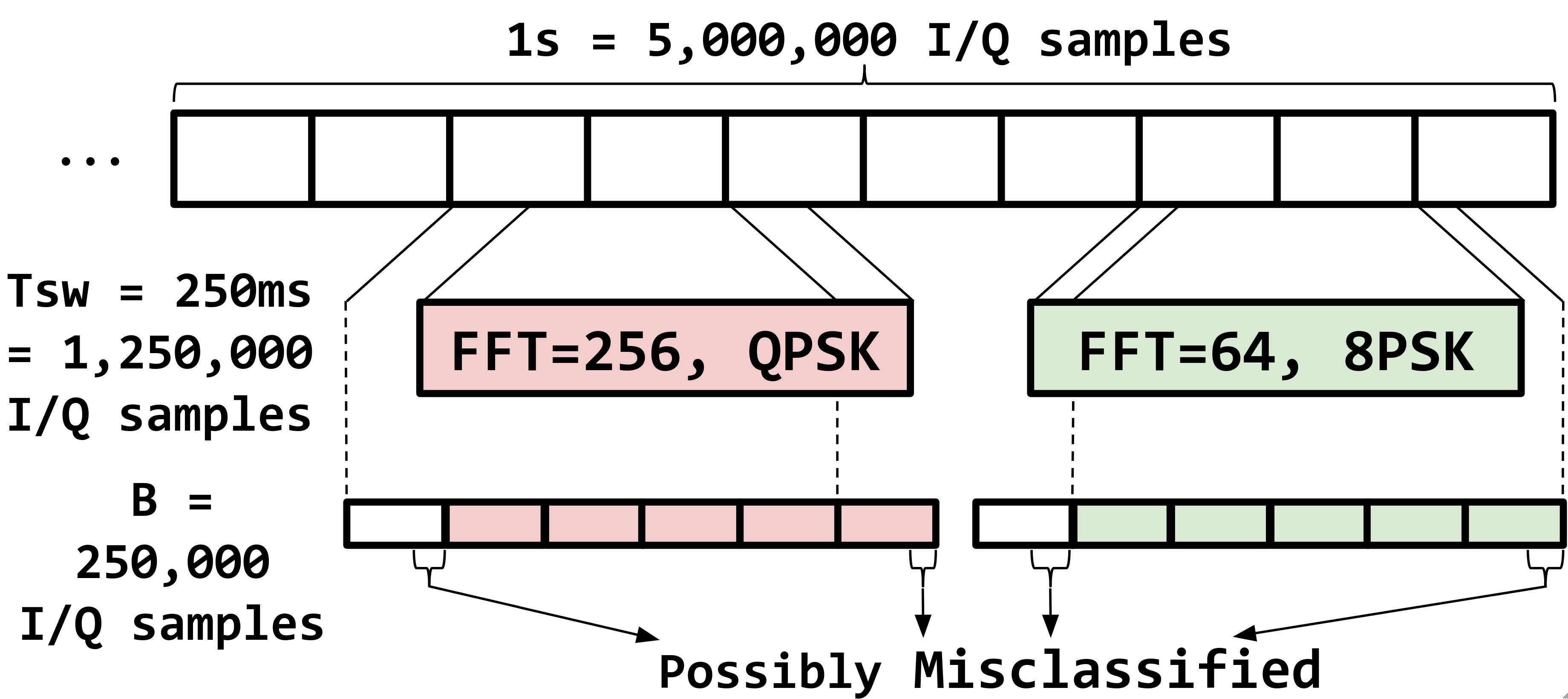}
    \caption{Switching time vs buffer size.\vspace{-0.2cm}}
    \label{fig:switching_time}
\end{figure}

The calculation above assumes that it only suffices to run the model once every switching time. However, since the receiver is not aligned with the switching time, we need to run \emph{RFNet} several times to obtain good performance. To help make this point, Figure \ref{fig:switching_time} shows an example where some of the I/Q samples may be misclassified due to misalignment. It can be shown that the amount of I/Q samples that can be misclassified due to misalignment is approximately $B/2$ on the average, assuming all classes are equally distributed (a more precise computation would involve knowing the number of classes and their distribution). In general, having a smaller buffer $B$ is desirable as it necessarily leads to less misclassifications. The trade-off here, however, is that a smaller $B$ implies running \textit{RFNet} more frequently, which leads to increased CPU usage in our current implementation. In our experiments, we run \textit{RFNet} 5 times per switching time, as shown in Figure \ref{fig:switching_time}. This compels us to further optimize RFNet's latency by employing FPGA parallelization techniques such as loop pipelining/unrolling.

% \begin{figure}[!h]
%     \centering
%     \includegraphics[width=\columnwidth]{figures/Layer.pdf}
%     \caption{Examples of loop unrolling/pipelining.\vspace{-0.2cm}}
%     \label{fig:pipeline}
% \end{figure}

%Figure \ref{fig:pipeline} shows an example of loop pipelining and unrolling, where a simple loop of three operations, \textit{i.e.}, OP1, OP2, and OP3, is executed twice. For simplicity, we assume that each operation takes one clock cycle to complete. With loop unrolling, we increase memory access to allow concurrent execution of the operation, bringing latency and interval to 1. With loop pipelining, the next OP1 operation is executed concurrently to the OP2 operation of the first loop iteration. This brings the interval to 1 clock cycle and the throughput to 1 output per clock cycle. Without unrolling/pipelining, the loop would need to wait 3 clock cycles to get a new input.

\begin{figure}[!h]
    \centering
    \includegraphics[width=\columnwidth]{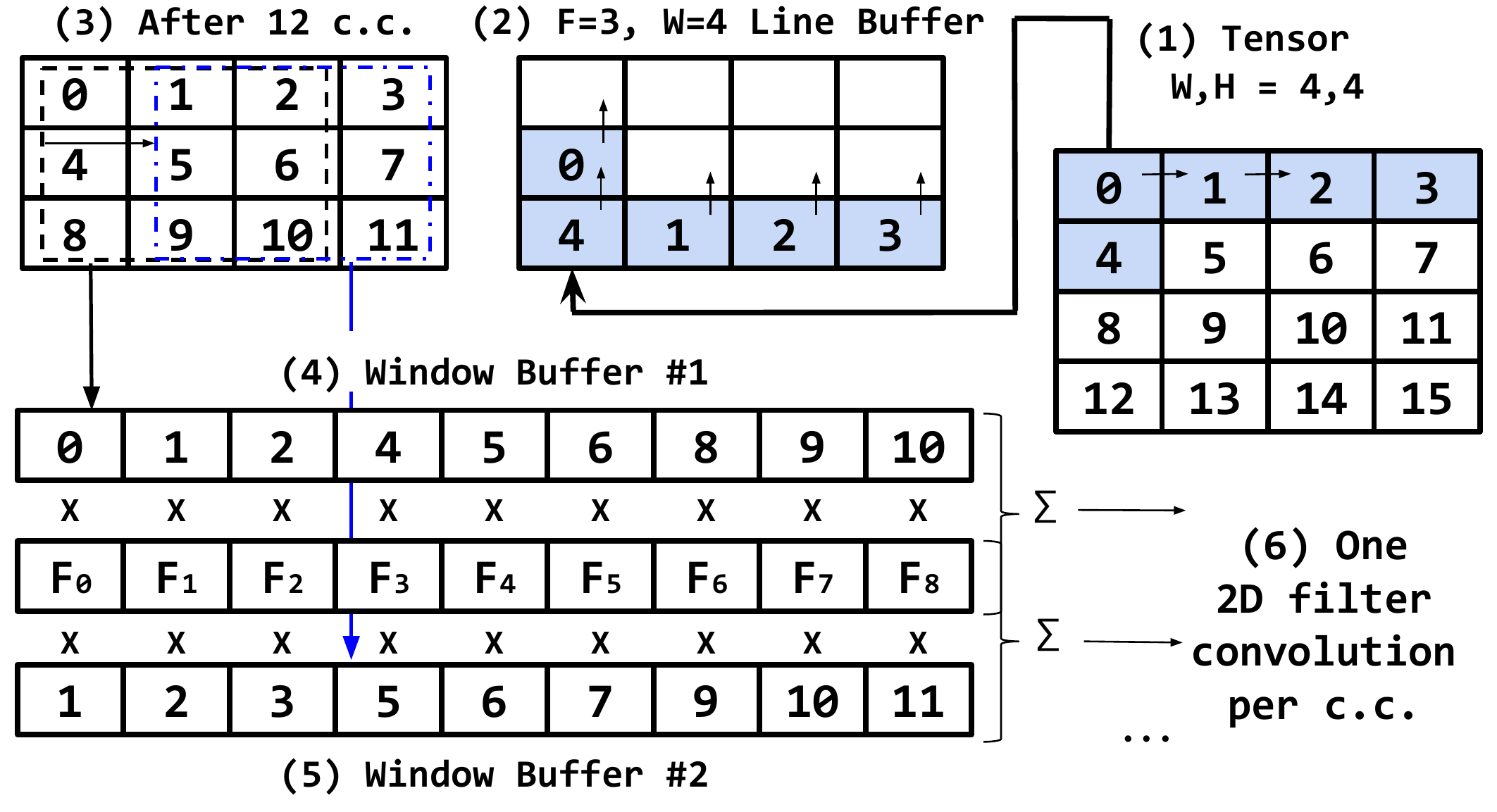}
    \caption{Example of line buffer and window buffer for 2D convolution operation. \vspace{-0.2cm}}
    \label{fig:line_window_buffer}
\end{figure}

\textbf{(2)}~\emph{Pipelined I/Q Convolution.}~Arguably, the most computationally expensive operation in \textit{RFNet} is the convolution between filters and input. To reduce the latency of this operation, \emph{we resort to a combination of pipelining/unrolling and line/window buffering}. Figure \ref{fig:line_window_buffer} shows an example of the combined functioning of the line and window buffers, where we set for simplicity $W,H = 4$ and $F=3$. In the example, I/Q samples are read one by one and inserted into the line buffer (step 1). During the insertion into the line buffer, elements are shifted vertically, while the data from the uppermost line is discarded (step 2), so as that after 12 clock cycles, the line buffer is filled (step 3). Since each line is located in a different memory location, the first window buffer can be filled by data from three line buffers in three clock cycles (step 4). The next window buffers simply discards the first column and reads the new column from the line buffers in one clock cycle. This, coupled with loop unrolling and pipelining, enables one filtering operation to be ready every clock cycle (step 5).

\subsection{PolymoRF: Linux Drivers}\label{sec:linux_drivers}

 The core challenge in designing drivers for embedded systems is that \textit{the same FPGA peripherals can change address assignment}. This would require us to re-compile the kernel every time an FPGA implementation uses different physical addresses, which is obviously not acceptable. To address this key issue, we had to resort to the device tree (DT) hardware description. The DT separates the kernel from the description of the peripherals, which are instead located in a separate binary file called the \textit{device tree blob} (DTB). This file is compiled from the \textit{device tree source} (DTS) file, and contains not only the customized \emph{PolymoRF} hardware addresses, but also every board-specific hardware peripheral information, such as the address of Ethernet, DMA, BRAMs, RF interface, and so on. 

\begin{figure}[!h]
    \centering
    \includegraphics[width=0.85\columnwidth]{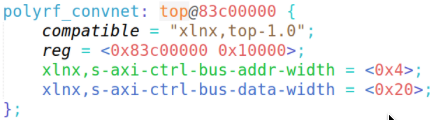}
    \caption{Device tree entry for the \textit{PolymoRF} ConvNet. \vspace{-0.2cm}}
    \label{fig:device_tree_entries}
\end{figure}

For \emph{PolymoRF}, we had to generate customized device-tree entries for the ConvNet, BRAM controllers, and AXI timer. Figure \ref{fig:device_tree_entries} shows the customized device tree entries in the DTS file for the input BRAM controller and RFNet (we omit the other entries due to space limitations). The most relevant entries are (i) the name of the peripheral, inserted before the '\{' character; (ii) the \textit{reg} entry, describing the starting physical address and the address space of the peripheral; and (iii) the \textit{compatible} entry, which defines the "programming model" of the device and allows the operating system to identify the corresponding device driver. Following the creation of the customized device-tree entries, we include these in the remainder portion of the global DTS file describing the ZC706 board, compile the DTB file and include it in the ZC706's SD card containing the operating system and the kernel image. %At bootstrap, the bootloader loads the kernel image and the DTB so that the kernel creates appropriate peripherals in the \texttt{/proc/device-tree/} directory. %Thus, our customized drivers can "bind" to the peripheral by using the corresponding device-tree entry, and access the AXI-Lite and AXI-Full memory space for read/write operations as it would access a file (also called "character device").

\section{Experimental Results}\label{sec:exp_results}

We first discuss details on our \emph{PolymoRF} prototype in Section \ref{sec:prototype_setup}, and then discuss the data collection and training process in Section \ref{sec:data_training}. We then investigate the performance of \emph{RFNet} in Section \ref{sec:learning_eval} on a single-carrier system. Then, we implement and test the throughput performance on a multi-carrier polymorphic OFDM system in Section \ref{sec:system-wide}. Finally, we report the latency and hardware performance of \emph{PolymoRF} in Section \ref{sec:hardware_eval}. \vspace{-0.1cm}

\subsection{Protoype and Experimental Setup}\label{sec:prototype_setup}

Our prototype is entirely based on off-the-shelf equipment. Specifically, we use a Xilinx Zynq-7000 XC7Z045-2FFG900C system-on-chip (SoC), which is a circuit integrating CPU, FPGA and I/O all on a single substrate \cite{Molanes-ieeetie2018}. We chose an SoC since it provides significant flexibility in the FPGA portion of the platform, thus allowing us to fully evaluate the trade-offs during system design. Moreover, the Zynq-7000 fully supports embedded Linux, which in effect makes the ZC706 a good prototype for a wireless platform. Our Zynq-7000 contains two ARM Cortex-A9 MPCore CPUs  and a Kintex-7 FPGA \cite{Zynq}, running on top of a Xilinx ZC706 evaluation board \cite{ZC706}. 

For both intra-FPGA and FPGA-CPU data exchange, we use the \textit{Advanced eXtensible Interface} (AXI) bus specification \cite{XilinxAXI}. In the AXI standard, data is exchanged during \textit{read} or \textit{write} \textit{transactions}. In each transaction, the \textit{AXI master} is charged with initiating the transfer; the \textit{AXI slave}, in turn, is tasked with responding to the AXI master with the result of the transaction (\textit{i.e.}, success/failure). An AXI master can have multiple AXI slaves, and \textit{vice versa}, according to the specific FPGA design. Multiple AXI masters/slaves can communicate with each other by using \textit{AXI interconnects}.  Specifically, \textit{AXI-Lite} is used for register access and configure the circuits inside the FPGA, while AXI-Stream is used to transport high-bandwidth streaming data inside the FPGA. AXI-Full is instead used by the CPU to read/write consecutive memory locations from/to the FPGA.

\begin{figure}[!h]
  \centering
  \includegraphics[width=1\linewidth]{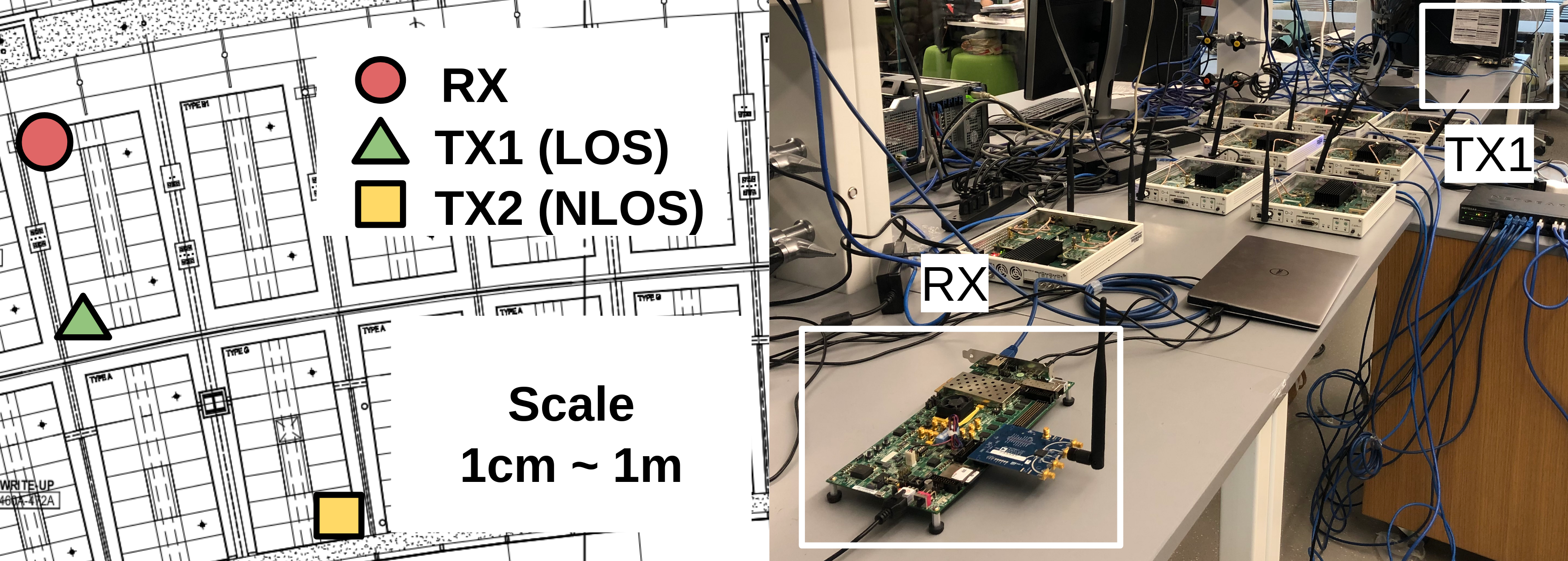}
   \caption{\label{fig:floor_plan}(left) Placement of the radios for experimental evaluation; (right) Experimental setting.  }\vspace{-0.2cm}
\end{figure}

To study \emph{PolymoRF} under realistic channel environments, we have used the experimental setup shown in Figure \ref{fig:floor_plan}. These scenarios investigate a line-of-sight (LOS) configuration where the transmitter is placed approximately 3m from the receiver, and a challenging non-line-of-sight (NLOS) channel condition where the transmitter is placed at 7m from the receiver and in the presence of several obstacles between them. Thus, the experiments were performed in a contested wireless environment with severe interference from nearby WiFi devices as well as multipath effect. 

\subsection{Data Collection and Training Process}\label{sec:data_training}

As far as the data collection and testing process is concerned, we first constructed a $\sim$10GB dataset by collecting waveform data in the line-of-sight (LOS) configuration, then used this data to train \emph{RFNet} through Keras. Then, we tested our models on live-collected data in both LOS and NLOS conditions. The transmitter radio used was a Zedboard equipped with an AD9361 as RF front-end and using Gnuradio for baseband processing. Waveforms were transmitted at center frequency of $2.432\  \textrm{GHz}$ (\textit{i.e.}, WiFi's channel 5).  %\textbf{To make our experiments replicable, we will make our $\sim$10GB dataset available to the community after publication.}

To train \textit{RFNet}, we use an $\ell_2$ regularization parameter $\lambda=0.0001$. We also use an Adam optimizer with a learning rate of $l = 10^{-4}$ and categorical cross-entropy as a loss function. All architectures are implemented in Python, on top of the Keras framework and with Tensorflow as the backend engine. %\textbf{The source code used to train the models is free and available to the community for download at \url{https://github.com/neu-spiral/RFMLS-NEU}.}

\subsection{Single-carrier Evaluation}\label{sec:learning_eval}

We consider the challenging problem of joint modulation and channel recognition in a single-carrier system where (i) modulation is chosen among BPSK, QPSK, 8PSK, 16-QAM, 32-QAM, and 64-QAM; (ii) spectrum is shifted of 0, 1~KHz and 2~KHz from its center frequency.  Due to space limitations, we only report results on the LOS scenario for the single-carrier scenario, and report in Section \ref{sec:system-wide} the performance of \textit{RFNet} on the NLOS scenario with the multi-carrier OFDM system.

\begin{figure}[!h] 
    \centering
    \includegraphics[width=0.35\columnwidth,angle=-90]{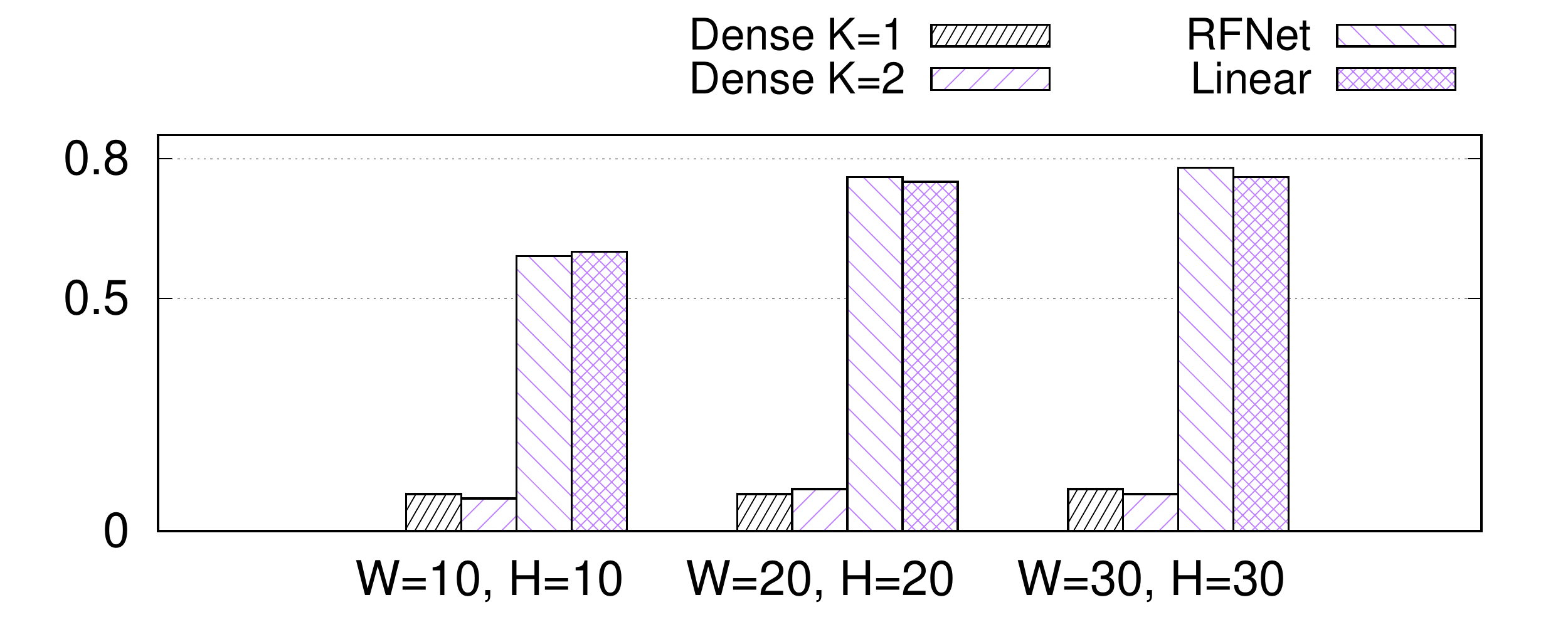}
    \caption{Comparison among \emph{RFNet}, \textit{Dense} and \textit{Linear} \cite{o2016convolutional}.}
    \label{fig:acc_comparison}
    \vspace{-0.3cm}
\end{figure}

\textbf{(1)}~\emph{Comparison with Existing Architectures.}~We compare \emph{RFNet} to \cite{o2016convolutional,OShea-ieeejstsp2018}, which is to the best of our knowledge  \cite{o2016convolutional,OShea-ieeejstsp2018} the current state of the art in RF waveform classification using ConvNets. This approach, called for simplicity \textit{Linear}, considers an input tensor of dimension $1 \times W \cdot H \times 2$ and convolutional layers with filters of dimension $1 \times F \times 2$. Thus, the filters in the first convolutional layer perform linear convolution over a set of $F$ consecutive I/Q samples. We attempted to train the architecture in \cite{OShea-ieeejstsp2018}, which has M = 7 convolutional layers with C=64 filters each and K=2 dense layers with 128 neurons each. However, due to its huge dimensions, we were not able to synthesize this architecture on our testbed. Therefore, we compared \textit{RFNet} with the architecture in \cite{o2016convolutional}, \textit{i.e.}, M=2 convolutional layers with C=256,80 and K=1 with 256 neurons. For fair comparison with \emph{Linear}, we selected the closest input size to ours (\textit{i.e.}, 1x128 vs 10x10, 1x400 vs 20x20, 1x900 vs 30x30).

Figure \ref{fig:acc_comparison}  shows the test-set accuracy obtained for a subset of the considered architectures, where \textit{RFNet} was trained with $M = 1$ convolutional layer with $C = 25$ filters, and no dense layer ($K = 0$). The obtained results indicate that traditional dense networks cannot recognize complex RF waveforms, as they attain slightly more accuracy (8\%) than the random-guess accuracy (5.5\%) -- regardless of the number of layers. This is because dense layers are not able to capture localized, small-scale I/Q variations in the input data, which is instead done by convolutional layers. Moreover, Figure \ref{fig:acc_comparison} indicates that \emph{RFNet} has similar accuracy as obtained by \textit{Linear}, despite using a much simpler architecture. This is due to the fundamental difference between how the convolutional layers in \emph{PolymoRF} and \emph{Linear} process I/Q samples.\smallskip

% \begin{figure}[!h]
%     \centering
%     \includegraphics[width=\columnwidth]{figures/Example_Filters_First_Layer.pdf}
%     \caption{First-layer filters in RFNet. \vspace{-0.2cm}}
%     \label{fig:filters_intuition}
% \end{figure}

% To confirm our intuition on how \emph{RFNet} learns transitions in the I/Q complex constellation plane, Figure \ref{fig:filters_intuition} shows the I/Q representation of two filters in the first convolutional layer of a model distinguishing between BPSK and QPSK. This experiment suggests that indeed \textit{RFNet} learns unique transitions in the I/Q plane, as the filters' output will be maximized for patterns that can only occur in certain modulation  -- respectively, transitions from the third to one quadrant and from second to fourth are more likely to occur in PSK modulation than in QAM modulations.

\begin{figure}[!h]
    \centering
    \includegraphics[width=0.35\columnwidth, angle=-90]{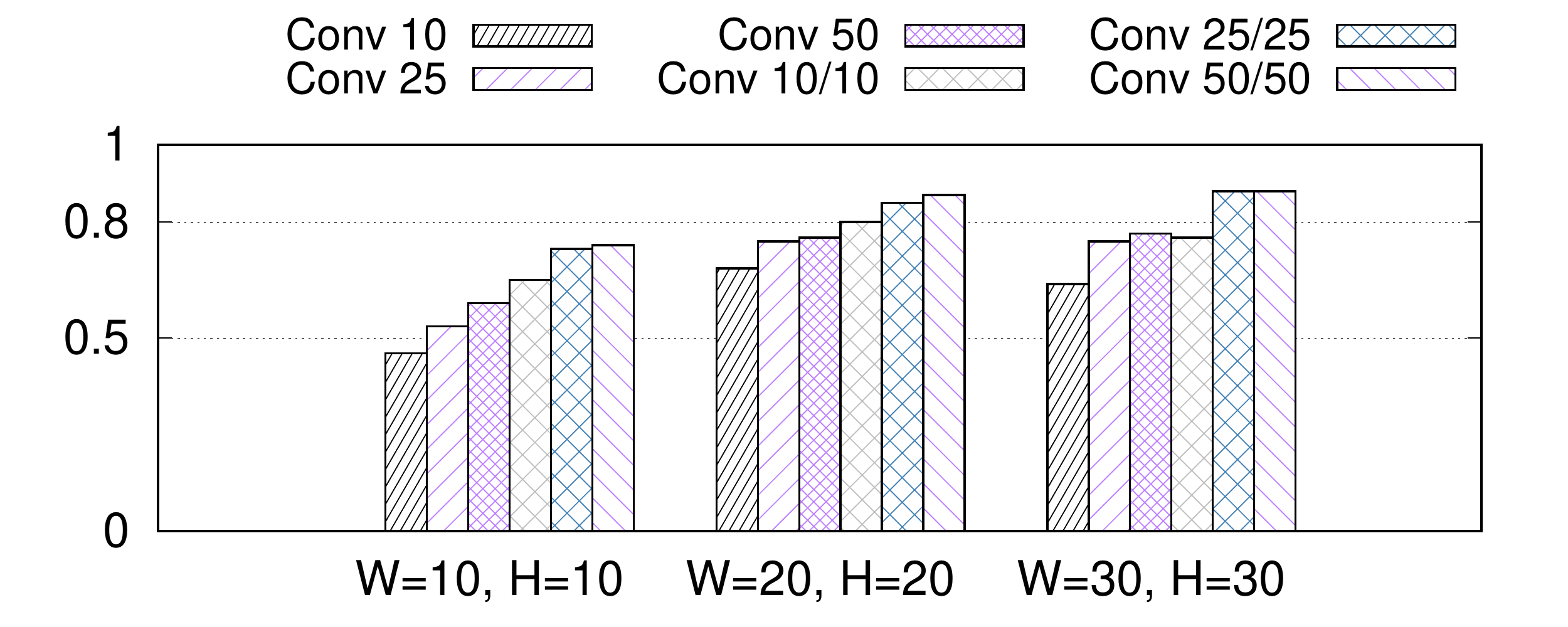}\\
    \includegraphics[width=\columnwidth]{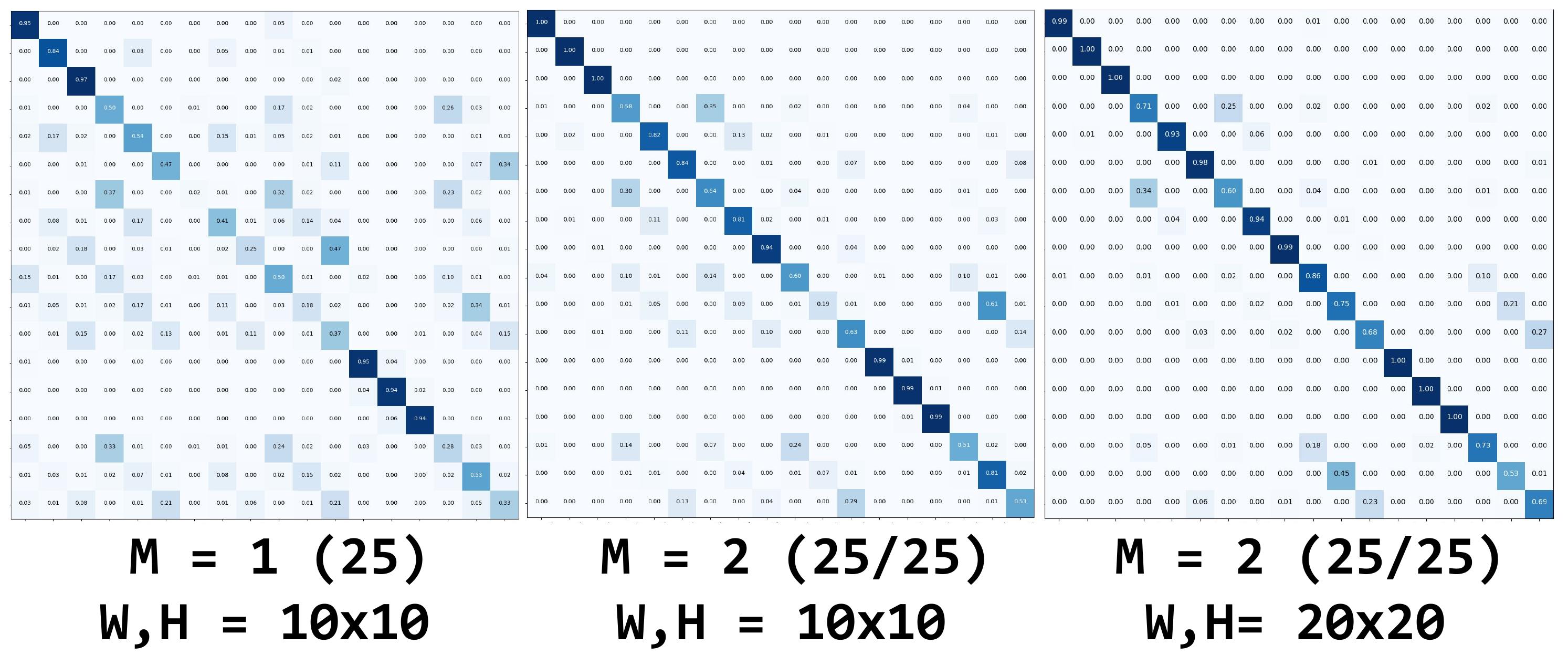}
    \caption{(top) Test-set classification accuracy vs input size $W$/$H$ vs $M$, with $K = 0$ (no dense layer); (bottom) Confusion matrices as function of $M$, $W$ and $H$.}
    \label{fig:acc_convlayers}
    \vspace{-0.2cm}
\end{figure}

\begin{figure}[!h]
    \centering
    \includegraphics[width=0.35\columnwidth, angle=-90]{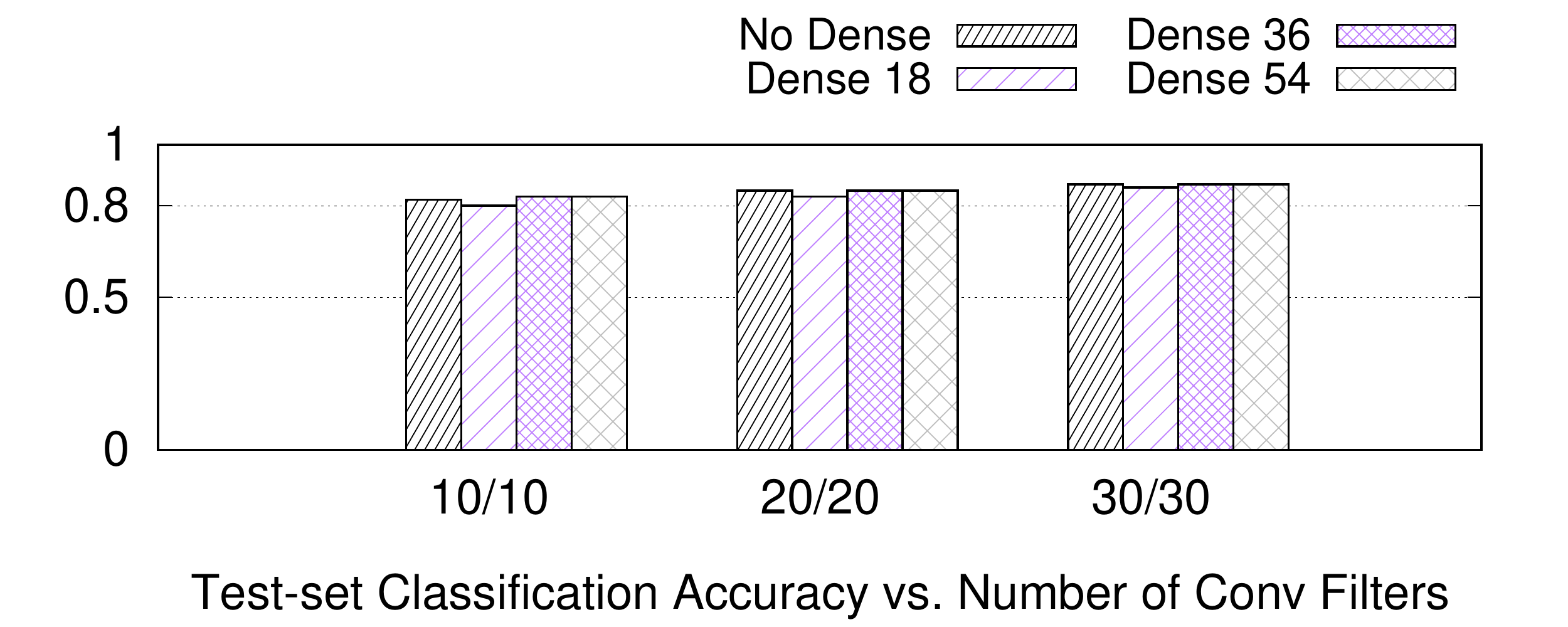}
    \caption{Accuracy vs number of filters vs dense layer size.}
    \label{fig:acc_denselayers}
    \vspace{-0.3cm}
\end{figure}

\textbf{(2)}~\emph{Hyper-parameter Evaluation.}~We study the impact of the number of convolutional layers $M$ and dense layers $K$, as well as the input size ($W$) and filter size ($F$) on the performance of \emph{RFNet}.  Figure \ref{fig:acc_convlayers} shows accuracy as a function of $W$ and $H$, for hyper-parameters M = {1, 2} and C = {10, 25, 50}. The results conclude that increasing $C$ does improve the performance but up to a certain extent. Indeed, we notice that switching to $C=50$ does not improve much the performance, especially when $M = 2$. This is because the number of distinguishing I/Q patterns is limited in number among different modulations, and thus the filters in excess end up learning similar patterns. Furthermore, increasing $W$ and $H$ increases accuracy significantly, since a larger input size allows to compensate for the adverse channels/noise conditions. Furthermore, Figure \ref{fig:acc_denselayers} illustrates the impact of $K$. Figure \ref{fig:acc_denselayers} suggests that the accuracy does not increase when adding a dense layer, regardless of its size, which indicates the correctness of our choice to exclude dense layers.

\begin{figure}[!h]
    \centering
    \includegraphics[width=0.35\columnwidth, angle=-90]{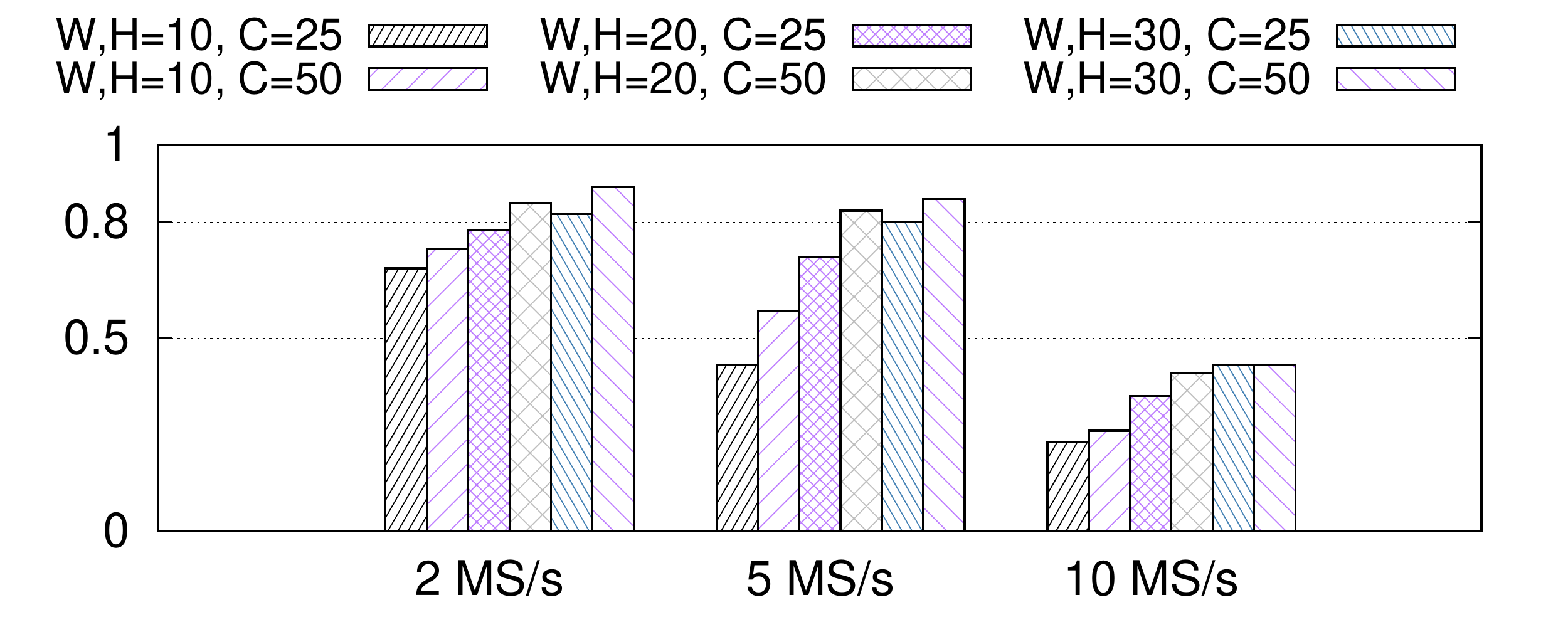}
    \caption{Accuracy vs transmitter's sampling rate.}
    \label{fig:acc_samplingrate}
    \vspace{-0.3cm}
\end{figure}

\begin{figure}[!h]
    \centering
        \includegraphics[width=0.82\columnwidth]{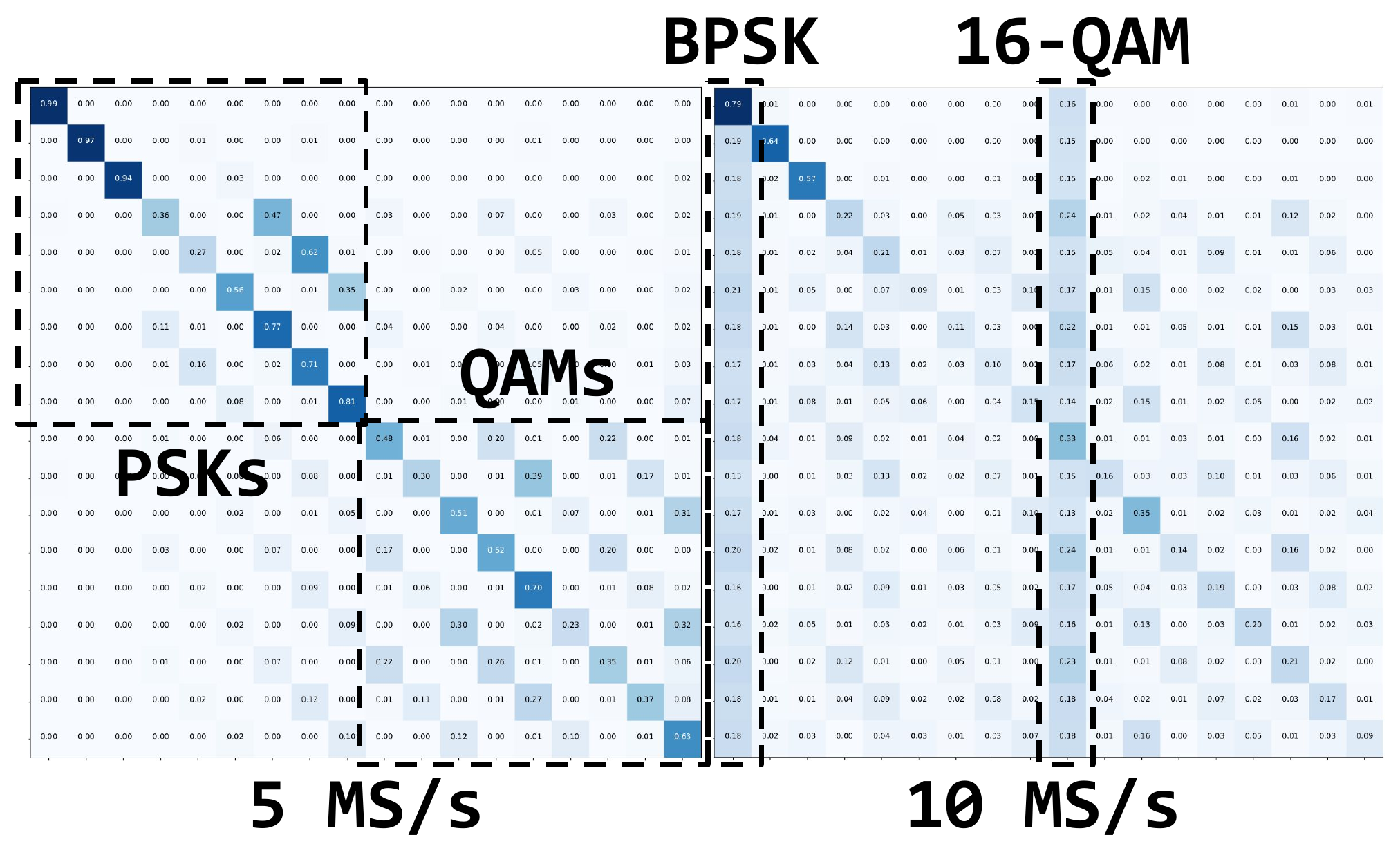}
    \caption{Confusion matrices for transmitter's sampling rate of 5 MS/s and 10 MS/s, W,H=10, C=50 model.}
    \label{fig:confusion_samplingrate}
    \vspace{-0.5cm}
\end{figure}

\textbf{(3)}~\emph{Impact of the Sampling Rate.}~We investigate the impact of the transmitter's sampling rate in Figure \ref{fig:acc_samplingrate}, where we show the classification accuracy  for different W, H and C values. We also show the confusion matrices\footnote{Class labels are ordered by modulation and frequency shift, \textit{i.e.}, from "BPSK, 0 KHz", "BPSK, 1 KHz", ... to "64-QAM, 2KHz".} for the W,H=10, C=50 architectures in Figure \ref{fig:confusion_samplingrate}. As expected, these results confirm that the performance of \emph{RFNet} decreases as the transmitter's sampling rate increases. This is because, as shown in Section \ref{sec:rf_images} \emph{RFNet} learns the I/Q transitions between the different modulations. Therefore, as the transmitter's sampling rate increases, the model will have fewer I/Q samples between the constellation points. Indeed, the confusion matrices show that with 5 MS/s the model becomes further confused with QAM constellations, and with 10 MS/s higher-order PSKs and QAMs ``collapse'' onto the lowest-order modulations. \smallskip

\textbf{(4)}~\emph{Remarks.}~The above results imply that oversampling the signal leads to a better modulation classification accuracy. However, we would like to point out that oversampling does not mean that the physical-layer has to process more data -- indeed, the extra samples can be dropped when going through the demodulation chain while the oversampled I/Q signal can be forwarded to \textit{RFNet} for classification. %, since it becomes challenging to distinguish among them.

\subsection{Multi-carrier Evaluation}\label{sec:system-wide}

We evaluated \emph{PolymoRF} on an OFDM system (in short, \emph{Poly-OFDM}) which supports 3 different FFT sizes (64, 128, 256) and 3 different symbol modulations in the FFT bins (BPSK, QPSK, 8PSK), creating in total a combination of 9 different parameter set which are switched pseudo-randomly by the transmitter.  A demo video where the transmitter switches FFT size every 0.5s is available at  \url{https://youtu.be/5vf_pb0nvKk}. In the following, we use the C=25,25, 20x20, pipelined \textit{RFNet} architecture, which presents latency of about 17ms (see Section \ref{sec:hardware_eval}). In these experiments, we set (i) the transmitter's sampling rate to 5M samples/sec; \emph{PolymoRF}'s buffer size $B$ to 250k I/Q samples; (iii) the switching time of the transmitter to 250ms. Thus,  \textit{RFNet} is run approximately 5 times during each switching time (see Section \ref{sec:par_strategies}). 

% \begin{figure}[!h]
%     \centering+
%         \includegraphics[width=0.9\columnwidth]{./figures/Confusion_Matrix_Experiments.pdf}
%     \caption{Confusion matrices for Poly-OFDM, (left) \emph{LOS} and (right) \emph{NLOS} scenarios.}
%     \label{fig:conf_PolymoRF}
%     \vspace{-0.3cm}
% \end{figure}

\begin{figure}[!h]
    \centering
    \includegraphics[width=0.35\columnwidth, angle=-90]{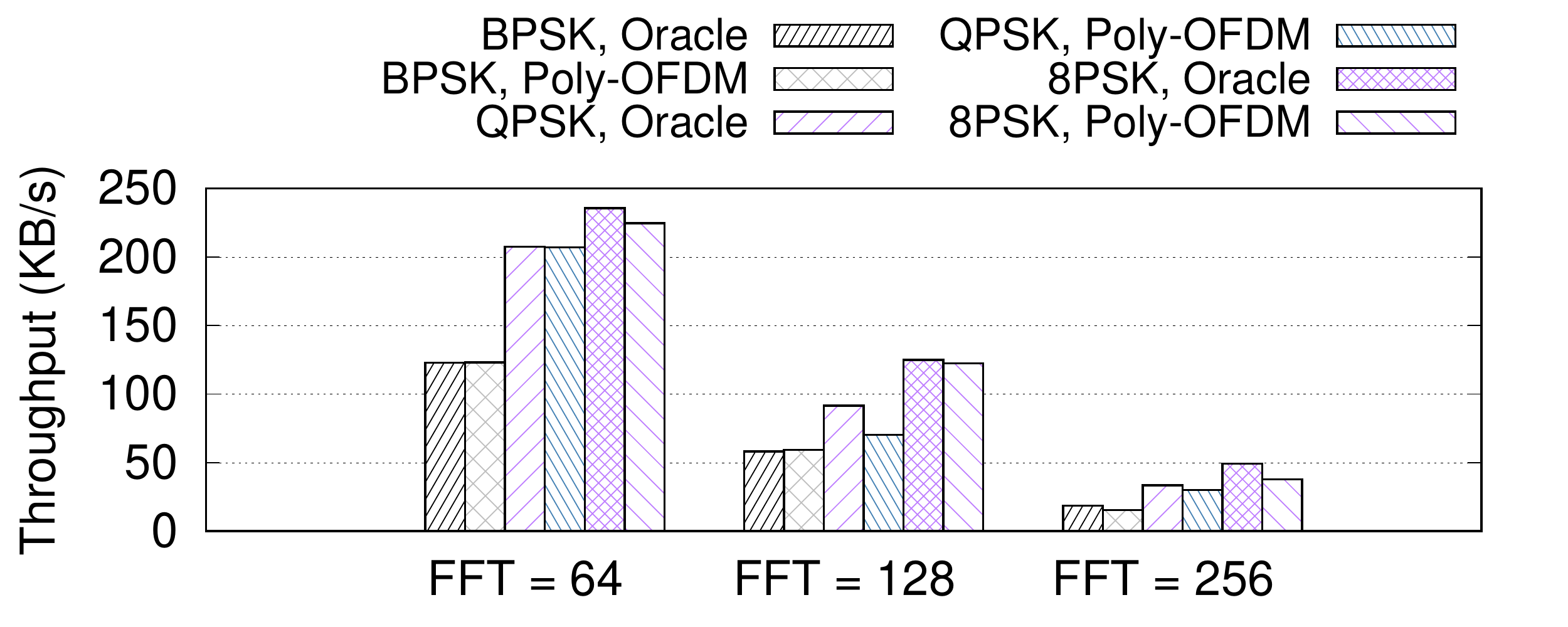}\\
        \includegraphics[width=0.38\columnwidth, angle=-90]{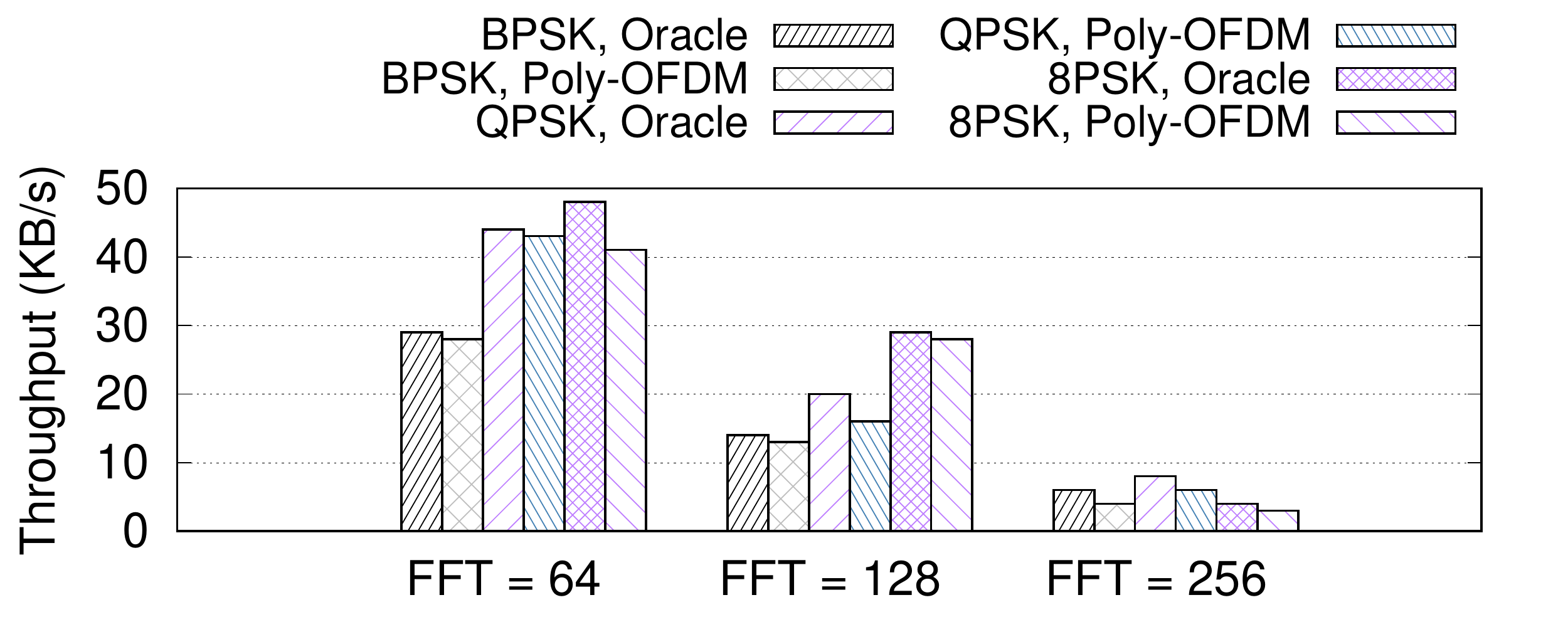}
    \caption{Comparison between \emph{Oracle} and \textit{Poly-OFDM},  (top) \emph{LOS} and (bottom) \emph{NLOS} scenarios.}
    \label{fig:throughput_polyofdm}
    \vspace{-0.2cm}
\end{figure}

The most critical aspect to be evaluated is how \emph{Poly-OFDM}, an inference-based system, compares with an ideal system that has perfect knowledge of the modulation and FFT size being used by the transmitter at each time, which we call for simplicity \emph{Oracle}. Although \emph{Oracle} cannot be implemented in practice, we believe this experiment is crucial to understand what is the throughput loss with respect to a system where the physical-layer configuration is known a priori. Figure \ref{fig:throughput_polyofdm}, where we show the comparison between \emph{Oracle} and \textit{Poly-OFDM} as a function of the FFT size and the symbol modulation. As we notice, the overall throughput results decrease in the NLOS scenario, which is expected given the impairments imposed by the challenging channel conditions.  On the other hand, the results in Figure \ref{fig:throughput_polyofdm} confirm that Poly-OFDM is able to obtain similar throughput performance with that of a traditional OFDM system, obtaining on the average 90\% and 87\% throughput of that of the traditional system.

\subsection{RFNet Latency Evaluation and Comparison}\label{sec:hardware_eval}

Table \ref{tab:hw_sw} compares latency, number of parameters, and BRAM occupation of \emph{RFNet} vs (i) a C++ implementation running in the CPU of our testbed; and (ii) existing work \cite{o2016convolutional}. We were not able to synthesize the 1x900 architecture, since it was too large for our hardware. As we can see, \emph{RFNet} is able to significantly reduce latency and memory occupation with respect to existing work. Indeed, we have a decrease of an order of magnitude in almost every considered scenario.  Table \ref{tab:latency} shows the comparison between the pipelined version of the ConvNet circuits and the CPU latency, as well as the look-up table (LUT) consumption increase with respect to the unpipelined version. Table \ref{tab:latency} concludes that on the average, our parallelization strategies bring close to $60\%$ and $100\%$ latency reduction with respect to the unoptimized and CPU versions, respectively, with a LUT utilization increase of about 7\% on the average.

\renewcommand{\arraystretch}{1.2}
\begin{table}[h!]
\centering
\begin{tabular}{|c|c|c|c|c|}
\hline
\cellcolor{blue!10}\textbf{Model} &  \cellcolor{green!10} \textbf{Input}  & \cellcolor{blue!10}\textbf{Latency}  & \cellcolor{green!10}\textbf{Params} & \cellcolor{blue!10}\textbf{BRAM}   \\ \hline
% \multirow{3}{*}{\parbox{1.1cm}{\centering RFNet \\ C=25}} & 10x10 & 2.918ms &   $\sim$11k & 1\%\\ \cline{2-5}
%           & 20x20 &  26.55ms &  $\sim$45k & 7\% \\ \cline{2-5}
%           & 30x30  & 93.35ms &   $\sim$81k & 15\% \\ \hline
\multirow{3}{*}{\parbox{1.1cm}{\centering RFNet \\ C=50}} & 10x10   & 5.835ms  & $\sim$23k & 3\%\\ \cline{2-5}
           & 20x20  & 53.11ms   & $\sim$90k & 14\% \\ \cline{2-5}
          & 30x30   & 233.3ms   & $\sim$203k & 29\% \\ \hline
\multirow{3}{*}{\parbox{1.1cm}{\centering RFNet \\ C=25,25}} & 10x10   & 6.704ms  &  $\sim$10k & 2\% \\ \cline{2-5}
           & 20x20  & 38.41ms  & $\sim$17k & 8\% \\ \cline{2-5}
          & 30x30   & 144.9ms   & $\sim$34k & 17\% \\ \hline
\multirow{3}{*}{\parbox{1.1cm}{\centering RFNet \\ C=50,50}} & 10x10   & 21.41ms   & $\sim$31k &4\% \\ \cline{2-5}
           & 20x20   & 100.3ms   & $\sim$40k & 16\% \\ \cline{2-5}
          & 30x30   & 336.9ms   & $\sim$81k & 34\% \\ \hline
\multirow{2}{*}{\parbox{1.1cm}{\centering Linear}} & 1x128 &   579.8ms  & $\sim$2M &16\% \\ \cline{2-5}
                    & 1x400  & 2,026ms & $\sim$8M & 64\% \\ \hline
\hline
\end{tabular}
\caption{Latency/hardware consumption evaluation.\vspace{-0.5cm}}
\label{tab:hw_sw}
\end{table}

\renewcommand{\arraystretch}{1.2}
\begin{table}[h!]
\centering

\begin{tabular}{|c|c|c|c|c|c|}
\hline
\cellcolor{blue!10}\textbf{Model} &  \cellcolor{green!10}\textbf{Input} &  \cellcolor{blue!10}\textbf{CPU} & \cellcolor{green!10}\textbf{Pipelined}  & \cellcolor{blue!10}\textbf{LUT}  \\ \hline
\multirow{3}{*}{\parbox{1.1cm}{\centering RFNet \\ C=25}} & 10x10 & 49.31ms & 1.19ms    & +3\%\\ \cline{2-5}
          & 20x20 & 478.4ms & 8.077ms     & +7\%  \\ \cline{2-5}
          & 30x30 & 1592ms & 25.54ms    & +9\% \\ \hline
\multirow{3}{*}{\parbox{1.1cm}{\centering RFNet \\ C=50}} & 10x10 & 106.4ms & 2.381ms    & +6\%\\ \cline{2-5}
          & 20x20 & 934.2ms & 16.15ms   & +12\%\\ \cline{2-5}
          & 30x30 & 3844ms & 63.81ms     & +20\% \\ \hline
\multirow{3}{*}{\parbox{1.1cm}{\centering RFNet \\ C=25,25}} & 10x10 & 122.1ms & 3.959ms     & +1\% \\ \cline{2-5}
          & 20x20 & 677.9ms & 16.29ms   & +4\% \\ \cline{2-5}
          & 30x30 & 2354ms& 49.57ms     & +7\% \\ \hline
\multirow{3}{*}{\parbox{1.1cm}{\centering RFNet \\ C=50,50}} & 10x10 & 363.9ms & 13.51ms   & +2\%  \\ \cline{2-5}
          & 20x20 & 1826ms & 48.87ms    & +7\% \\ \cline{2-5}
          & 30x30 & 5728ms & 131.7ms    & +11\% \\ \hline
\hline
\end{tabular}\vspace{0.1cm}
\caption{Pipelined vs CPU latency.}\vspace{-0.8cm}
\label{tab:latency}
\end{table}

\begin{figure}[!h]
  \centering
  \includegraphics[width=0.9\linewidth]{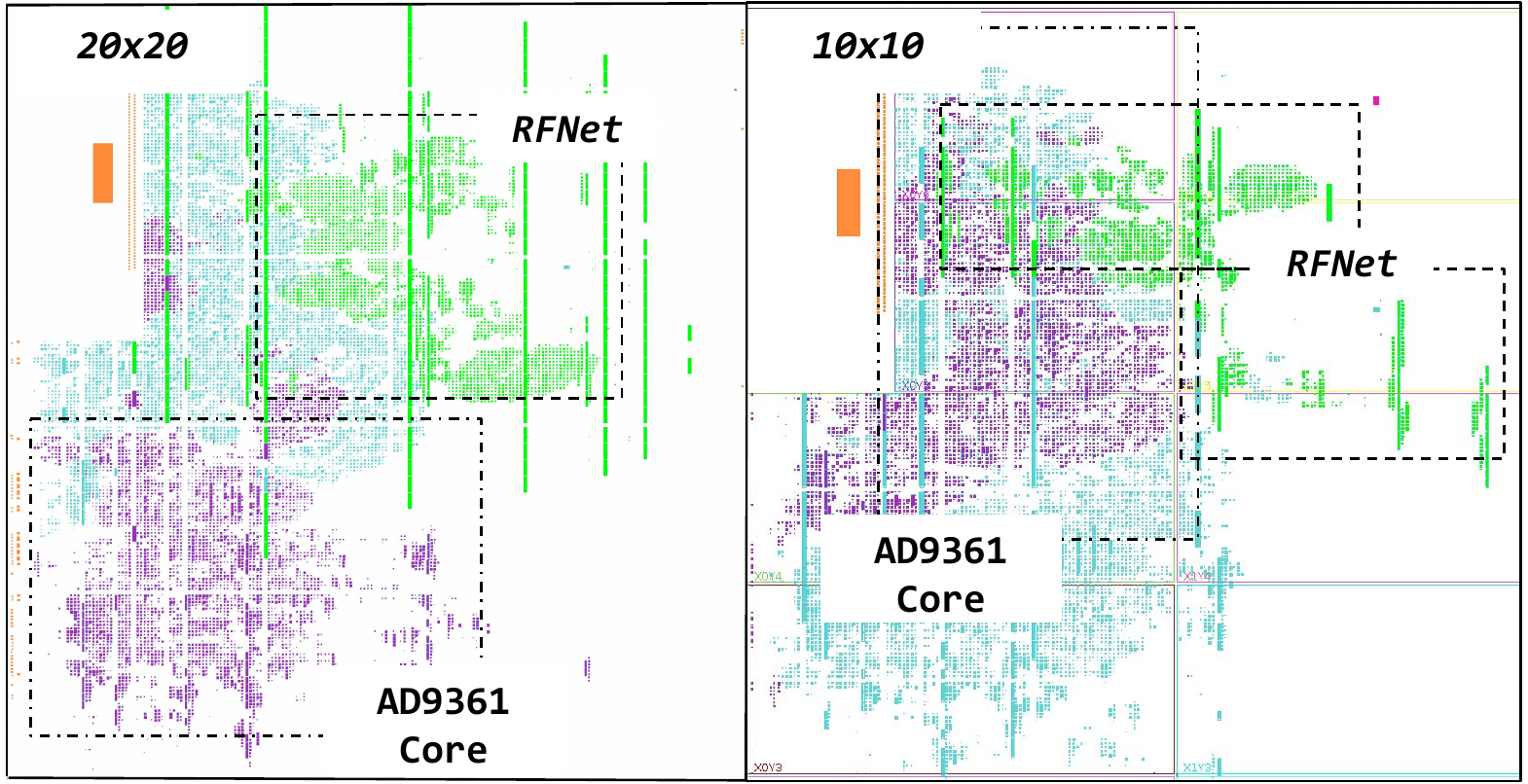}
  \caption{\label{fig:fpga}\textit{PolymoRF} FPGA implementations.}\vspace{-0.4cm}
\end{figure}

To give the reader a perspective of the amount of resources consumed on the FPGA, Figure \ref{fig:fpga} shows the FPGA implementation of respectively 10x10 and 20x20 \textit{RFNet} model, both pipelined and with C=25,25 architecture, where we highlight and color the resource consumption of \textit{RFNet} with respect to the AD9361 circuitry. Figure \ref{fig:fpga} indicates that the resource consumption of the \emph{RFNet} circuit is significantly lesser than the AD9361 one in the 10x10 case, and becomes comparable with the 20x20 architecture. In any case, the overall resource consumption of our FPGA designs is about 50\% of the total FPGA resources.

\section{Related Work and Conclusions}

Learning-based radios are envisioned to be able to automatically infer the current spectrum status in terms of occupancy \cite{subramaniam2015spectrum}, interference \cite{chen2016survey} and malicious activities \cite{jin2018specguard}.  Most of the existing work is based on low-dimensional machine learning \cite{Pawar-ieeetifs2011,Shi-ieeetcomm2012,Ghodeshar-icscn2015,xiong2019robust}, which requires the cumbersome manual extraction of very complex, ad hoc features from the waveforms. For this reason, deep learning has been proposed as a viable alternative to traditional learning techniques \cite{Mao-ieeecomm2018}. The key problem of RF modulation recognition through deep learning has been extensively investigated \cite{OShea-ieeejstsp2018,o2017introduction,wang2017deep,West-dyspan2017,Kulin-ieeeaccess2018,Karra-ieeedyspan2017}. The seminal work by  O'Shea \emph{et al.}  \cite{OShea-ieeejstsp2018} and Karra \emph{et al.} \cite{Karra-ieeedyspan2017} proposed ConvNets-based to address the issue. However, the authors do not address the issue of what to do with the inferred RF information. Conversely, Kulin \emph{et al.} present in \cite{Kulin-ieeeaccess2018} a framework for end-to-end wireless deep learning, where a use case on dynamic spectrum access is provided. The above work proposes models leveraging a significant number of parameters, thus ultimately not applicable to real-time RF settings. Recently, \cite{Restuccia-infocom2019} has demonstrated the need for real-time hardware-based RF deep learning. However, the main limitation of \cite{Restuccia-infocom2019} is that it focus on the learning aspect only, ultimately  not addressing the problem of connecting real-time inference with receiver reconfigurability.

\emph{Summary and Ongoing Work.}~This paper has proposed \emph{PolymoRF}, a prototype that can be reused to develop and test novel polymorphic wireless communication systems. One of the key insights brought by our experimental evaluation is that the RF channel may impact the performance of \emph{RFNet} to a significant extent. To this end, we can (i) train different learning models for different channels and reconfigure the weights of \emph{RFNet} in the FPGA accordingly; (ii) apply small, controlled modifications to the RF signal at the transmitter's side to compensate the current RF channel condition. Another core aspect is the impact of polymorphism on the effectiveness of smart jamming attacks. We are conscious that the above issues are definitely worth investigating, however they deserve separate papers and are the subject of our ongoing work.

\section{Acknowledgements}

This  work  is  supported  in part by  the  Office  of  Naval  Research (ONR)  under contracts N00014-18-9-0001. The views and conclusions contained herein are those of the authors and should not be interpreted as necessarily representing the official policies or endorsements of the ONR or the U.S. Government.

\footnotesize
\bibliographystyle{IEEEtran}
\bibliography{main} 

\end{document}